\documentclass[preprint,notoc,float]{JHEP3}

\usepackage{epsfig}
\usepackage{axodraw}

\def\eslt{\not\!\!{E_T}}

\def\to{\rightarrow}

\def\pT{$p_{T}$ }
\def\bi{\begin{itemize}}
\def\ei{\end{itemize}}
\def\be{\begin{equation}}
\def\ee{\end{equation}}
\def\bea{\begin{eqnarray}}
\def\eea{\end{eqnarray}}

\def\ttb{t \overline{t} }

\def\X{\times}
\def\emis{\not\hskip-5truedd E_{T} }
\def\LQ{{L\!\!Q}}

\title{Leptoquark Single and Pair production at LHC with  CalcHEP/CompHEP  in the  complete model}

\author{
Alexander Belyaev$^1$, 
Claude Leroy$^2$,
Rashid Mehdiyev{$^2$}%
\footnote{On leave of absence from Institute
of Physics, ANAS, Baku, AZ-370143, Azerbaijan},
Alexander Pukhov$^3$
\\
$^1$ Department of Physics and Astronomy, 
East Lansing, 48824, USA\\
E-mail: \email{belyaev@pa.msu.edu.edu}\\
$^2$ Universit\'e de Montr\'eal, D\'epartement de Physique,
		  Montr\'eal, QC, H3C 3J7, Canada\\
E-mail: \email{Leroy@lps.umontreal.ca, RMehdi@lps.umontreal.ca}\\ 
$^3$ 
Skobeltsyn Institute of Nuclear Physics,
Moscow State University, Moscow 119992,
Russia\\
E-mail: \email{pukhov@sinp.msu.ru}
}

\preprint{\vbox{\hbox{MSU-HEP-070204}}} 

\abstract{
We study combined leptoquark ($\LQ$) single and pair production at LHC
at the level of detector simulation.
A set of kinematical cuts  to maximize 
significance for combined signal events has been  worked out.

It was  shown that 
combination of signatures from  $\LQ$ single and pair production
not only significantly increases the LHC reach, but also 
allows us to give the correct  signal interpretation.
In particular, it was found that  the LHC has potential to discover
$\LQ$ with a mass up to     $1.2$~TeV  and $1.5$~TeV 
for the case of scalar and vector $\LQ$, respectively,
and $\LQ$ single production contributes about 30-50\% to the total signal 
rate for $\LQ-l-q$ coupling, taken equal to the electromagnetic coupling.

This work is based on an implementation 
of the most general form of scalar and vector $\LQ$ interactions 
with quarks and gluons into CalcHEP/CompHEP packages.
This implementation, which authors made publicly available,
was one the most important aspects of the  study.

}


\begin{document}
\section{Introduction}

Boson fields mediating lepton-quark interactions  naturally appear in
various extensions of the Standard Model which is known to be theoretically incomplete.
Leptoquarks appear in the framework of Grand Unified theories (GUT)
where quarks and leptons are unified in one matter multiplet~\cite{guts},
in the SUSY models with R-parity violation
(in this case the mediator of the lepton-quark interaction is a squark or a slepton),
as well as in composite models of leptons and quarks~\cite{Schrempp:1984nj}.

In general, boson fields give rise to violation of the baryon and lepton numbers 
(leading to fast proton decay)
and flavor changing neutral current(FCNC) processes
which are strongly constrained by experiment. 
In principle, those bosons should be  heavy~($\sim M_{Planck}\sim 10^{19}$~GeV)
to suppress these unwelcome processes.
On the other hand, fast proton decay and FCNC problems
can be avoided  if the boson mass even is of the order of the electroweak~(EW) scale.
In the case of leptoquarks~($\LQ$), which exclusively induce lepton-quark
interactions, the problems above are solved:
1)  the fast proton decay
problem is absent, since
leptoquarks conserve the baryon and lepton numbers (in the case of squarks mediating 
lepton-quarks interactions, in R-parity violating SUSY models, the proton decay is  
preserved when only  {\it one,}  lepton {\it or }  baryon number is conserved);  
2) the FCNC problem is also absent under the assumption
of  flavor diagonal form of {\it leptoquark-lepton-quark} ($\LQ-l-q$) interactions. 

Numerous experimental $\LQ$ searches at HERA~(e.g.~\cite{lq-hera})
and at the Tevatron~(for  recent results see, e.g.~\cite{lq-tev}
and references therein) gave no positive results so far and provided
only limits on $\LQ$ masses and $\LQ$ couplings to leptons and quarks.
Eventually, the CERN Large Hadron Collider~(LHC) will be able to extend significantly
the reach for $\LQ$ masses and couplings~(see, e.g. 
\cite{Dion:1997jw,Eboli:1997fb,CiezaMontalvo:1998sk,%
Belyaev:1998ki,Abdullin:1999im}). In~\cite{Kramer:1997hh,Kramer:2004df}
it was shown that Next-to-Leading-Order (NLO) corrections to $\LQ$ pair production
are positive and non-negligible ( up to 20\% at the Tevatron and up to 90\%
at the LHC) and lead to further extension of the collider reach in $\LQ$ searches.
Leptoquarks can be produced not only in pairs (from gluon splitting)
but also as a single particle in association with 
lepton~\cite{Hewett:1987yg,Eboli:1987vb,Dion:1996rv}.
In~\cite{Eboli:1999ye} authors noticed the importance
of the combination of $\LQ$ single and pair production analysis for the Tevatron 
collider, which partially motivated the present study.

In this article, we perform a new detailed  study of  $\LQ$ production and decay 
at  LHC.
There are several motivations for this work.
Firstly, we stress the importance of the study of {\it combined} 
$\LQ$ single and pair production,
since both contribute to the same  signature at the detector simulation level.
Therefore, combination of signatures from $\LQ$ single and pair production
not only significantly increases the LHC reach, but also 
allows us to give the correct  signal interpretation.
Secondly, we study the complete set of $\LQ$ interactions
including {\it scalar and vector} $\LQ$ production and decay.
Finally, one should stress that this study is based on the
implementation of the complete $\LQ$ model including 
{\it the most general form of scalar and vector $\LQ$ interactions 
with quarks and gauge bosons
(including gluons)} into CalcHEP/CompHEP packages~\cite{calc,comp}.
This is one of the most important aspects of this work.
Such an implementation allowed  us to study  the effect of vector $\LQ$ interactions
with gluons via anomalous couplings of the most general form.

This paper is organized as follows. In section 2, we describe
the general effective Lagrangian used in our study,
as well as the implementation of this Lagrangian into the CalcHEP/CompHEP
software package. In section 3, we present signal rates for $\LQ$ single and pair 
production at the LHC  for the case of scalar and vector $\LQ$.
In section 4, we perform an analysis of signal versus background
at the detector level.
Finally, in section 5, we draw conclusions on  LHC
potential for leptoquarks search.

\section{$\LQ$ model and its implementation into CalcHEP/CompHEP packages}

\subsection{The model}

Following~\cite{Buchmuller:1986zs,Blumlein:1992ej},
we use  an effective Lagrangian with the most general dimensionless,
$SU(3)\times SU(2)\times U(1)$ invariant couplings
of scalar and vector leptoquarks to leptons and quarks with lepton and baryon number 
conservation: 
\begin{equation}
\label{eq-lq-general}
{\cal L}={\cal L}^f_{|F|=0} +{\cal L}^f_{|F|=2} +{\cal L}^V
\end{equation}
Lagrangian ${\cal L}^f_{|F|=0,2}$ describes Yukawa type interactions of $\LQ$ with
leptons and quarks ($\LQ-l-q$), changing the fermion number $F$ by 0 or 2,
respectively, where $F=3B+L$, \ $B$ is the baryon number and $L$ is the
lepton number. ${\cal L}^f_{|F|=0,2}$ conserves the baryon and lepton numbers
and has a flavor diagonal form:
\begin{eqnarray}
\label{eq:e-q-lq0}
{\cal L}^f_{|F|=0}
&=&
(h_{2L}\bar{u}_R\ell_L+h_{2R}\bar{q}_L i \tau_2 e_R)R_2 
+ 
\tilde{h}_{2L}\bar{d}_R\ell_L\tilde{R}_2 
\nonumber\\
&+&
(h_{1L}\bar{q}_L\ell_L+h_{1R}\bar{d}_R \gamma^\mu e_R)U_{1\mu} 
\nonumber\\
&+&
\tilde{h}_{1R}\bar{u}_R\gamma^\mu e_R\tilde U_{1\mu}
+      h_{3L}\bar{q}_L \vec{\bf\tau}\gamma^\mu\ell_L\vec{U}_{3\mu}+h.c.,
\\
\nonumber
\\
\label{eq:e-q-lq2}
{\cal L}^f_{|F|=2}
&=&
(g_{1L}\bar{q}_L^c i \tau_2 \ell_L +g_{1R}\bar{u}_R^c e_R)S_1 
\nonumber\\
&+&
\tilde{g}_{1R} \bar{d}_R^c e_R\tilde{S}_1+ 
 g_{3L}\bar{q}_L^c i\tau_2\vec{\tau}\ell_L\vec{S}_3
\nonumber\\
&+&
(g_{2L}\bar{d}_R^c\gamma^\mu \ell_L  +  g_{2R}\bar{q}_L^c\gamma^\mu e_R)V_{2\mu}
\nonumber\\
&+&
\tilde{g}_{2} \bar{u}_R^c \gamma^\mu\ell_L\tilde{V}_{2\mu}+h.c.
\end{eqnarray} 
where $\tau_i$ are the Pauli matrices,
$q_L$  and $\ell_L$ are $SU(2)_L$ quark and lepton doublets, respectively
and $u_R$, $d_R$, and $e_R$ are corresponding singlet fields;
charged conjugated fields are denoted by $f^c=C\bar{f}^T$.
We follow the $\LQ$ classification from~\cite{Blumlein:1992ej}. 
Table~\ref{tab:lq-table} summarizes the  complete set of  scalar and vector
 $\LQ$ fields appearing in Eq.~(\ref{eq:e-q-lq0})-(\ref{eq:e-q-lq2}):
 $S_1$, $\tilde{S}_1$,  $\vec{S}_3$,   $R_2$,  $\tilde{R}_2$,
and
$V_2^\mu$, $\tilde{V}_2^\mu$,  
$U_1^\mu$, $\tilde{U}_1^\mu$, $\vec{U}_3$,  respectively.
Scalar  ($S_1,\tilde{S}_1,\vec{S}_3,R_2,\tilde{R}_2$)
and vector ($V_2^\mu, \tilde{V}_2^\mu, U_1^\mu, \tilde{U}_1^\mu, \vec{U}_3$)
$\LQ$s have Yukawa-type  couplings to  quarks and leptons 
denoted by 
($g_{1(L,R)},\tilde{g}_{1R},g_{3L},h_{2(L,R)},\tilde{h}_{2L}$)
and
($g_{2(L,R)},\tilde{g}_{2R},h_{1(L,R)},\tilde{h}_{1R},h_{3L}$)
respectively, in Eq.~(\ref{eq:e-q-lq0})-(\ref{eq:e-q-lq2}).
Couplings appearing in the Feynman
rules  of  $\LQ-q-l$~($\LQ-q-\nu$)  interactions
(which we denote by $\lambda_L(lq)$, $\lambda_R(lq)$ and $\lambda_L(\nu q)$)
are trivial linear combinations of $g,h,\tilde{g},\tilde{h}$
couplings presented in Table~\ref{tab:lq-table}.
The table  also presents a notation for  $\LQ$ names
in the model realized in CalcHEP/CompHEP packages which we describe below.

\TABLE{
\begin{tabular}{|c|c|r|c|r|r|r|r|r|ccc|}
\hline
&&&&&&&&&\multicolumn{3}{c|}{CalcHEP/}\\
$\LQ$($\Phi$)         & Spin & F & Color   &  $T_3$&$Q_{em}$&$\lambda_L(lq)$& $\lambda_R(lq)$& $\lambda_L(\nu q) $&\multicolumn{3}{c|}{CompHEP}\\
&&&&&&&&&\multicolumn{3}{c|}{notation($\Phi/\overline{\Phi}$)}\\
\hline
\hline
$S_1$  	           & 0    & -2&$\bar{3}$&$ 0   $&$+1/3 $  &$g_{1L}         $&$g_{1R}         $&$-g_{1L}$        &S1&/&s1 \\\hline
$ \tilde{S}_1$     & 0	  & -2&$\bar{3}$&$ 0   $&$+4/3 $  &$0              $&$\tilde{g}_{1R} $&$0      $        &ST&/&st \\\hline
                   &	  &   &        	&$ +1  $&$+4/3 $  &$-\sqrt{2}g_{3L}$&$0 	     $&$0      $        &SP&/&sp \\
$   \vec{S}_3$     & 0    & -2&$\bar{3}$&$ 0   $&$+1/3 $  &$-g_{3L}        $&$0 	     $&$-g_{3L}$	&S0&/&s0 \\
                   &	  &   &        	&$ -1  $&$-2/3 $  &$0              $&$0 	     $&$\sqrt{2}g_{3L} $&SM&/&sm \\\hline
                   &      &   &        	&$ 1/2 $&$+5/3 $  &$h_{2L}         $&$h_{2R}         $&$0      $        &rp&/&RP \\
$       {R}_2$     & 0	  &  0&$    {3}$&$     $&$     $  &		    &		      &		        &  & &   \\
                   &	  &   &        	&$ -1/2$&$+2/3 $  &$0              $&$-h_{2R}        $&$h_{2L}$	        &rm&/&RM \\\hline
                   &      &   &        	&$ +1/2$&$+2/3 $  &$\tilde{h}_{2L} $&$0              $&$0     $	        &tp&/&TP \\
$ \tilde{R}_2$     & 0	  &  0&$    {3}$&$     $&$     $  &		    &		      &		        &  & &   \\
                   &	  &   & 	&$ -1/2$&$-1/3 $  &$0$  	    &$0              $&$\tilde{h}_{2L} $&tm&/&TM \\\hline\hline
                   &	  &   & 	&$ +1/2$&$+4/3 $  &${g}_{2L}       $&$g_{2R}         $&$0$	        &VP&/&vp \\
$      {V}_{2\mu} $& 1	  & -2&$\bar{3}$&$     $&$     $  &		    &		      &		        &  & &   \\
                   &	  &   & 	&$ -1/2$&$+1/3 $  &$0              $&$g_{2R}         $&$g_{2L}$	        &VM&/&vm \\\hline
                   &	  &   & 	&$ +1/2$&$+1/3 $  &$\tilde{g}_{2L} $&$0              $&$0     $	        &WP&/&wp \\
$\tilde{V}_{2\mu} $& 1	  & -2&$\bar{3}$&$     $&$     $  &		    &		      &		        &  & &   \\
                   &	  &   & 	&$ -1/2$&$-2/3 $  &$0              $&$0              $&$\tilde{g}_{2L} $&WM&/&wm \\\hline
$      {U}_{1\mu} $& 1	  &  0&$    {3}$&$  0  $&$+2/3 $  &$h_{1L}         $&$h_{1R}         $&$h_{1L}$	        &u1&/&U1 \\\hline
$\tilde{U}_{1\mu} $& 1	  &  0&$    {3}$&$  0  $&$+5/3 $  &$0              $&$\tilde{h}_{1R} $&$0     $ 	&ut&/&UT \\\hline
                   &      &   & 	&$ +1  $&$+5/3 $  &$\sqrt{2}h_{3L} $&$0              $&$0     $	        &up&/&UP \\
$   \vec{U}_{3\mu}$& 1	  &  0&$    {3}$&$  0  $&$+2/3 $  &$-h_{3L}        $&$0              $&$h_{3L}$	        &u0&/&U0 \\
                   &	  &   & 	&$ -1  $&$-1/3 $  &$0              $&$0              $&$\sqrt{2}h_{3L} $&um&/&UM \\\hline\hline
\end{tabular}
\caption{\label{tab:lq-table}  
Quantum numbers for the  complete set of  scalar and vector
 $\LQ$ fields appearing in Eq.~\ref{eq:e-q-lq0},\ref{eq:e-q-lq2}:
\ \  $S_1$, $\tilde{S}_1$,  $\vec{S}_3$,   $R_2$,  $\tilde{R}_2$,
and
$V_2^\mu$, $\tilde{V}_2^\mu$,  
$U_1^\mu$, $\tilde{U}_1^\mu$, $\vec{U}_3$,  respectively~\cite{Blumlein:1992ej}.
Also,  $\LQ$-lepton-quark couplings [$\lambda_L(lq)$, $\lambda_R(lq)$ and $\lambda_L(\nu q)$]
and notation for $\LQ$ names
for the model realized in CalcHEP/CompHEP are presented.
The particle-antiparticle convention is defined as:
$\overline{\Phi}_{F=2}\rightarrow lq$ and $\overline{\Phi}_{F=0}\rightarrow l\bar{q}$. 
}
}

The $\LQ$ interactions with gauge bosons are described by ${\cal L}^V$,
obey the 
$SU(3)_c\times SU(2)\times U(1)_Y$ Standard Model symmetry.

The $\LQ-gluon$ interactions are 
described by the Lagrangian of the most general 
form~\cite{Blumlein:1996qp}
for the scalar and vector $\LQ$ interactions, ${\cal L}_S^g$ and ${\cal L}_V^g$, 
respectively:
\begin{eqnarray}
\label{eq-lq-glu-scal}
{\cal L}_S^g
 &=& \sum_{scalars} \left [
\left (D^{\mu}_{ij} \Phi^j \right )^{\dagger}
                                 \left (D_{\mu}^{ik} \Phi_k \right )
 - M_S^2 \Phi^{i \dagger} \Phi_i \right ],
\\
\label{eq-lq-glu-vec}
{\cal L}_V^g
&=&
\sum_{vectors} \left \{ -\frac{1}{2} V^{i \dagger}_{\mu \nu}
V^{\mu \nu}_i + M_V^2 \Phi_{\mu}^{i \dagger} \Phi^{\mu}_i 
-
ig_s \left [ (1 - \kappa_G)
\Phi_{\mu}^{i \dagger}
t^a_{ij}
\Phi_{\nu}^j
{\cal G}^{\mu \nu}_a
+ \frac{\lambda_G}{M_V^2} 
V^{i\dagger}_{\sigma \mu}t^a_{ij}
V_{\nu}^{j \mu} {\cal G}^{\nu \sigma}_a \right ] \right \}.
\nonumber\\
\end{eqnarray}
Here,
$g_s$ denotes the  strong coupling
constant, $t_a$ are the generators of $SU(3)_c$,
$M_S$($M_V$)
are
the scalar(vector) leptoquark masses, while  $\kappa_G$ and
$\lambda_G$ are the anomalous couplings related to the
anomalous magnetic and quadrupole moments of vector $\LQ$~\cite{Blumlein:1996qp}.
Fields $\Phi$ and $\Phi^\mu$ represent 
scalar and vector leptoquarks, respectively.
The field strength tensors of the  gluon and vector
leptoquark fields are:
\begin{eqnarray}
{\cal G}_{\mu \nu}^a  &=& \partial_{\mu} G_{\nu}^a
 - \partial_{\nu}
G_{\mu}^a + g_s f^{abc} G_{\mu b} G_{\nu c},
 \nonumber\\
V_{\mu \nu}^{i}
 &=& D_{\mu}^{ik}
 \Phi_{\nu k} - D_{\nu}^{ik} \Phi_{\mu k},
\end{eqnarray}
with the covariant derivative given by
\begin{equation}
D_{\mu}^{ij} = \partial_{\mu} \delta^{ij} - i g_s
t_a^{ij} G^a_{\mu}.
\end{equation}

We omit here the analogous  $U(1)_Y\times SU(2)$ piece of gauge interactions 
since it is not relevant for our study of $\LQ$ production at LHC,
where (gauge boson -- $\LQ$) interactions are  eventually gluon dominant.

\subsection{$\LQ$ color factorization 
and model implementation into CalcHEP/CompHEP}

The appearance of vertex with 4 color particles can not be straightforwardly implemented
into CalcHEP/CompHEP packages.
The idea is to split 4-color interactions into 3-color vertices
via the introduction of auxiliary ghosts fields.

It is well known that  for $G\bar{q}q$ and $GGG$ QCD vertices 
their color
structure  is  factorisable. For  calculation of  Feynman 
diagrams  with such  vertices, where  color indexes can be convolved separately, 
an elegant technique is presented in \cite{Cvitanovic}.  
However, for $4G$ vertex 
such a factorization is absent. 
To have  color factorization  in  the case of $4G$ vertex one can split 
this vertex into $3G$  vertex by means of the auxiliary
tensor field  ${G_t}_{\mu\nu}$\cite{comp}. 
This field should have the point-like propagator 
\begin{equation}
<0|T[{G_t}^{\mu_1\nu_1}_{\alpha_1}(p_1), \; {G_t}^{\mu_2\nu_2}_{\alpha_2}(p_2)]|0> = \frac{1}{(2
\pi)^4i} \;
\delta(p_1+p_2) \,\delta_{\alpha_1\alpha_2} g^{\mu_1\mu_2}\, g^{\nu_1\nu_2}\;.
\label{tensor_propagator}
\end{equation}
and interact with gluons according to 
\begin{equation}
\label{tGG-lgrng}
 S_{G_tGG}=\frac{i\, g}{\sqrt{2}}\int  f^{\alpha}_{\beta\gamma}
{{G_t}_{\alpha}}^{\mu\nu}(x)  G^{\beta}_{\mu}(x)G^{\gamma}_{\nu}(x) d^4x
\end{equation}
Graphically, it can be represented as 
\FIGURE[h]{
\begin{picture}(80,80)(0,0)
\Text(10,70)[r]{$G_1$}
\Text(10,10)[r]{$G_2$}
\Text(70,70)[l]{$G_3$}
\Text(70,10)[l]{$G_4$}

\DashLine(10,70)(70,10){3.0} 
\DashLine(70,70)(10,10){3.0}
\Vertex(40,40){2} 
\end{picture} 
\begin{picture}(10,80)(0,0)
\Text(5,40)[]{=}
\end{picture}
\begin{picture}(80,80)(0,0)
\Text(10,70)[r]{$G_1$}
\Text(10,10)[r]{$G_2$}
\Text(70,70)[l]{$G_3$}
\Text(70,10)[l]{$G_4$}

\DashLine(10,70)(10,10){3.0} 
\DashLine(70,70)(70,10){3.0} 
\DashLine(10,40)(70,40){3.0} 

\Text(40,42.0)[b]{$G_t$}
\Vertex(10,40){2}
\Vertex(70,40){2}
\end{picture} 
\begin{picture}(10,80)(0,0)
\Text(5,40)[]{+}
\end{picture}
\begin{picture}(80,80)(0,0)
\Text(10,70)[r]{$G_1$}
\Text(10,10)[r]{$G_2$}
\Text(70,70)[l]{$G_3$}
\Text(70,10)[l]{$G_4$}
\DashLine(10,70)(70,70){3.0} 
\DashLine(10,10)(70,10){3.0} 
\DashLine(40,10)(40,70){3.0} 
\Text(38,40.0)[r]{$G_t$}
\Vertex(40,10){2}
\Vertex(40,70){2}
\end{picture} 
\begin{picture}(10,80)(0,0)
\Text(5,40)[]{+}
\end{picture}
\begin{picture}(80,80)(0,0)
\Text(10,70)[r]{$G_1$}
\Text(10,10)[r]{$G_2$}
\Text(70,70)[l]{$G_3$}
\Text(70,10)[l]{$G_4$}
\DashLine(10,70)(70,10){3.0} 
\DashLine(70,70)(10,10){3.0} 
\DashLine(55,55)(55,25){3.0} 
\Text(57,40.0)[l]{$G_t$}
\Vertex(55,55){2}
\Vertex(55,25){2}
\end{picture} 
\\
\caption{\label{4Gsplitt} Splitting of a four-gluon vertex into three-gluon vertex.}}

This trick was successfully implemented  in  CompHEP/CalcHEP  
packages  for QCD Lagrangian. However in the case  of the interaction of  vector
leptoquarks  with gluon~\cite{Blumlein:1996qp} one needs to  develop this
technique further and find a general prescription for  
splitting of the arbitrary color structure.
Such a  prescription is described below.
We consider separately cases of real (e.g. gluon) and complex 
(e.g. leptoquark) vector fields.

In the case of a {\bf real vector field} $\Phi_\mu$ interacting with a gluon field 
$G_\mu$, the Lagrangian can be presented as 
\begin{equation}
L= -\frac{1}{4}|\partial_{\mu}\Phi_{\nu} - \partial_{\nu}\Phi_{\mu} +\Upsilon_{\mu\nu}|^2
+ Int(\Upsilon(\Phi,G), \Phi,...)
\end{equation}
where  $\Upsilon_{\mu\nu}=g[\Phi_{\mu},G_{\nu}]$, 
$Int$ represents  terms of gluon interaction with color matter realized
via general derivative or the strength tensor $F_{\mu\nu}$. The color index
of the gluon fields is omitted  to simplify the expressions.

 Now,   we  replace  $\Upsilon_{\mu\nu}$   by an 
 auxiliary  field $T_{\mu\nu}$ and  confine of
$\Upsilon=T$ by means of a Lagrange multiplier $t^{\mu\nu}$. We apply this procedure 
for  the
$Int$  and $\Upsilon(\Phi,G)^2$ terms, keeping the linear $\Upsilon$ term  as it is:
\begin{eqnarray}
L&=& -\frac{1}{4}|\partial_{\mu}\Phi_{\nu} - \partial_{\nu}\Phi_{\mu}|^2
  -\frac{1}{2} (\partial_{\mu}\Phi_{\nu} -
\partial_{\nu}\Phi_{\mu})\Upsilon(\Phi,G)^{\mu\nu}
\nonumber\\
&&  -\frac{1}{4}T_{\mu\nu}T^{\mu\nu}
   +\frac{1}{2}t_{\mu\nu}(T^{\mu\nu}-\Upsilon(\Phi,G)^{\mu\nu}) 
\nonumber\\
&&+Int(\Upsilon\to X,\Phi,...)
\end{eqnarray}
Then, we perform the following substitution of variables
\begin{eqnarray}
{\Phi_T}_{\mu\nu}=(T_{\mu\nu} - t_{\mu\nu})/\sqrt{2} \ \ \  \mbox{and} \ \ \ 
{\Phi_t}_{\mu\nu}=i\cdot t_{\mu\nu}/\sqrt{2}
\end{eqnarray}
in order to obtain standard normalized quadratic forms for the auxiliary fields:
\begin{eqnarray}
L&=& -\frac{1}{4}|\partial_{\mu}\Phi_{\nu} - \partial_{\nu}\Phi_{\mu}|^2
  -\frac{1}{2} (\partial_{\mu}\Phi_{\nu} -
\partial_{\nu}G_{\mu})\Upsilon(\Phi,G)^{\mu\nu}\nonumber\\
&&  -\frac{1}{2}{\Phi_T}_{\mu\nu}{\Phi_T}^{\mu\nu}
    -\frac{1}{2}{\Phi_t}_{\mu\nu}{\Phi_t}^{\mu\nu}
   +{i\over \sqrt{2}}{\Phi_t}\Upsilon(\Phi,G)^{\mu\nu} 
\nonumber\\
&& +Int(\Upsilon\to\sqrt{2}(\Phi_T-i \cdot \Phi_t),G,...)
\end{eqnarray}
In the case of QCD interaction with scalar and fermion fields, the $Int()$
term does not contain $\Upsilon$. Thus, the auxiliary field $T$ is free and can be
omitted. Other terms produce the  $3-gluon$ interaction, the point-like propagator
(\ref{tensor_propagator}), and the interaction of {\it just one} auxiliary
tensor  field $t$ with gluons ~(\ref{tGG-lgrng}).

Lagrangian with a {\bf complex vector} field $\Phi$  also can be represented in
a color-factored form using the same trick. In  a general case,  it is given as: 
\begin{equation}
L= -\frac{1}{2}|\partial_{\mu}\Phi_{\nu} - \partial_{\nu}\Phi_{\mu} +
  \Upsilon(\Phi,G)_{\mu\nu}|^2
+ Int(\Upsilon(\Phi,G), \Upsilon(\Phi,G)^*, \Phi, \Phi^*, ...)
\end{equation}
Following the procedure adopted with real vector field, we introduce two auxiliary fields
$T_{\mu\nu}$ and $t_{\mu\nu}$  which are now complex. In
terms of these fields, the Lagrangian becomes: 
\begin{eqnarray}
L&=& -\frac{1}{2}|\partial_{\mu}\Phi_{\nu} - \partial_{\nu}\Phi_{\mu}|^2
  -\frac{1}{2} (\partial_{\mu}\Phi_{\nu}^* -
\partial_{\nu}\Phi_{\mu}^*)\Upsilon(\Phi,G)^{\mu\nu}
  -\frac{1}{2} \Upsilon(\Phi,G)^*_{\mu\nu} (\partial^{\mu}V^{\nu} -
\partial^{\nu}V^{\mu})      					
\nonumber\\
&&-\frac{1}{2}T_{\mu\nu}^*T^{\mu\nu}
   +\frac{1}{2}t_{\mu\nu}^*(T^{\mu\nu}-\Upsilon(\Phi,G)^{\mu\nu})
+\frac{1}{2}(T_{\mu\nu}^*-\Upsilon(\Phi,G)_{\mu\nu}^*)t^{\mu\nu}            
\nonumber\\
&&+Int(\Upsilon\to T,\Upsilon(\Phi,G)^*\to T^* ...)
\end{eqnarray}

To obtain a standard normalized quadratic form of  auxiliary fields, one
should perform the following variable substitutions:
\begin{eqnarray}
{\Phi_T}_{\mu\nu}=(T_{\mu\nu} - t_{\mu\nu})/\sqrt{2}\;,\;\;
{\Phi_T}_{\mu\nu}^*=(T_{\mu\nu}^*- t_{\mu\nu}^*)/\sqrt{2} 			       \nonumber\\
{\Phi_t}_{\mu\nu}=i\cdot t_{\mu\nu}/\sqrt{2}\;, \;\;but!\;\; {\Phi_t}^*=i\cdot
t_{\mu\nu}^*/\sqrt{2}\;; 
\end{eqnarray}

Note that this transformation is realized through a {\it non-analytic} manner
which   is legal for  functional integrals.
Now, the Lagrangian in color-factorized form can be given as
\begin{eqnarray}								
L&=& -\frac{1}{2}|\partial_{\mu}\Phi_{\nu} - \partial_{\nu}\Phi_{\mu}|^2
  -\frac{1}{2} (\partial_{\mu}\Phi_{\nu}^* -
\partial_{\nu}\Phi_{\mu}^*)\Upsilon(\Phi,G)^{\mu\nu}
  -\frac{1}{2} \Upsilon(\Phi,G)^*_{\mu\nu} (\partial^{\mu}\Phi^{\nu} -
\partial^{\nu}\Phi^{\mu})     							  
\nonumber\\
 &&
  -|\Phi_T|^2 -|\Phi_t|^2
+{i\over \sqrt{2}} {\Phi_t}_{\mu\nu}^*\Upsilon(\Phi,G)^{\mu\nu}
+{i\over \sqrt{2}} \Upsilon(\Phi,G)_{\mu\nu}^*{\Phi_t}^{\mu\nu} 				
\nonumber\\
&&
+Int(\Upsilon\to\sqrt{2}(\Phi_T-i\cdot\Phi_t), \Upsilon(\Phi,G)^*\to\sqrt{2}(\Phi_T^*-i\cdot \Phi_t^*) ...)
\end{eqnarray}
In particular, for the case of Lagrangian~(\ref{eq-lq-glu-vec}) 
one has
\begin{eqnarray}								
&&\Upsilon(\Phi,G)_{\mu\nu}   = -ig(G_\mu\Phi_\nu-G_\nu\Phi_\mu)\\
&&\Upsilon(\Phi,G)_{\mu\nu}^* =  ig(G_\mu\Phi_\nu^\dagger-G_\nu\Phi_\mu^\dagger)\\
&&Int(\Upsilon\to\sqrt{2}(\Phi_T-i\cdot\Phi_t), 
    \Upsilon(\Phi,G)^*\to\sqrt{2}(\Phi_T^*-i\cdot \Phi_t^*) ...)=\nonumber\\
&&
    -ig(1-\kappa_G)\Phi_\mu t^a\Phi_\nu
    [\partial_\mu G_\nu^a-\partial_\nu G_\mu^a+
    \sqrt{2}({\Phi_t}_{\mu\nu}^a-i {\Phi_T}_{\mu\nu}^a)]\nonumber\\
&&    -ig{\lambda_G\over M_{\Phi}^2 }
   [{\partial_\sigma}{\Phi^\dagger}_\mu-{\partial_\mu}{\Phi^\dagger}_\sigma
     + \sqrt{2}({\Phi_T}_{\sigma\mu}^*-i {\Phi_t}_{\sigma\mu}^*)]\times\nonumber\\
&&   [{\partial_\nu}{\Phi}^\mu-{\partial^\mu}{\Phi}_\nu
    + \sqrt{2}({\Phi_T}_\nu^\mu-i {\Phi_t}_\nu^\mu)]\times
     [({\partial^\nu}G^\sigma-{\partial^\sigma}G^\nu
     + \sqrt{2}({G_T}^{\nu\sigma}-i {G_t}^{\nu\sigma})]
\end{eqnarray}
In the case of a {\bf complex scalar}  field  
one should introduce two auxiliary {\bf vector} fields
-- $V_\mu$ and $v_\mu$ (analogous to $T_{\mu\nu}$
and $t_{\mu\nu}$ tensors).
The Lagrangian with the  complex scalar field $\Phi$  can be
represented  as:  
\begin{equation}
L= |\partial_{\mu}\Phi +\Upsilon(\Phi,G)_{\mu}|^2
+ Int(\Upsilon(\Phi,G), \Upsilon(\Phi,G)^*, \Phi, \Phi^*, ...)
\end{equation}
In terms of these fields, in exact analogy with the vector leptoquark field,
we have 
\begin{eqnarray}
L&=& |\partial_{\mu}\Phi|^2
  +\partial_{\mu}\Phi^*\Upsilon(\Phi,G)^{\mu}
  +\Upsilon(\Phi,G)^*_{\mu} \partial^{\mu}\Phi 
\nonumber\\
&&+V_{\mu}^*V^{\mu}
   - v_{\mu}^*[V^{\mu}-\Upsilon(\Phi,G)^{\mu}]
   -[V_{\mu}^*-\Upsilon(\Phi,G)_{\mu}^*]v^{\mu}            
\nonumber\\
&&+Int(\Upsilon\to V,\Upsilon(\Phi,G)^*\to V^* ...)
\end{eqnarray}
with the same variable substitutions:
\begin{eqnarray}
{\Phi_V}_{\mu}  =(V_{\mu}  - v_{\mu})/\sqrt{2}& &
{\Phi_V}_{\mu}^*=(V_{\mu}^*- v_{\mu}^*)/\sqrt{2} 			       
\nonumber\\
{\Phi_v}_{\mu}=i\cdot v_{\mu}/\sqrt{2} &\ \  but! \ \  & 
{\Phi_v}^*=i\cdot v_{\mu\nu}^*/\sqrt{2} 
\end{eqnarray}

For Lagrangian~(\ref{eq-lq-glu-scal}) 
one has:
\begin{eqnarray}								
&&\Upsilon(\Phi,G)_{\mu}   = -ig G_\mu\Phi\\
&&\Upsilon(\Phi,G)_{\mu}^* =  ig G_\mu\Phi\\
&&Int(\Upsilon\to\sqrt{2}(\Phi_V-i\cdot\Phi_v), 
    \Upsilon(\Phi,G)^*\to\sqrt{2}(\Phi_V^*-i\cdot \Phi_v^*) ...)=0.
\end{eqnarray}
\begin{table}
\footnotesize{
\begin{verbatim}
P1   |P2   |P3   |  Factor              | Lorentz Part
---------------------------------------------------------------------------------------------
vm   |VM   |G    |  GG/MVM^2            |MVM^2*((1-KG)*(m2.p3*m1.m3-m1.p3*m2.m3)
                                        |     +(p2.m1*m2.m3-p1.m2*m1.m3+(p1-p2).m3*m1.m2))
                                        |      +LG*( p3.m1*(p1.m2*p2.m3-p1.p2*m2.m3)
                                        |           -p3.m2*(p1.m3*p2.m1-p1.p2*m1.m3)
                                        |           +p2.p3*(p1.m3*m1.m2-p1.m2*m1.m3)
                                        |          -p1.p3*(p2.m3*m1.m2-p2.m1*m2.m3))                                                                                                                                                                                                                                                                                                                                                                                                      
vm   |VM.t |G    |  GG/MVM^2/Sqrt2      |MVM^2*( m1.m2*m3.M2-m1.M2*m3.m2)
                                        |+2*LG*( p3.m2*(p1.m3*m1.M2-p1.M2*m1.m3)
                                        |       +m3.m2*(p1.M2*p3.m1-p1.p3*m1.M2))                                                                                                                                                                                                                                                                                                                                                                                                                                                                                                                      
VM   |vm.t |G    | -GG/MVM^2/Sqrt2      |MVM^2*(m1.m2*m3.M2-m1.M2*m3.m2)
                                        |+2*LG*( p3.m2*(p1.m3*m1.M2-p1.M2*m1.m3)
                                        |        +m3.m2*(p1.M2*p3.m1-p1.p3*m1.M2))                                                                                                                                                                                                                                                                                                                                                                                                                                                                                                                      
vm   |VM.T |G    | i*GG/MVM^2*Sqrt2     |LG*( p3.m2*(p1.m3*m1.M2-p1.M2*m1.m3)
                                        |    +m3.m2*(p1.M2*p3.m1-p1.p3*m1.M2))                                                                                                                                                                                                                                                                                                                                                                                                                                                                                                                                                             
VM   |vm.T |G    |-i*GG/MVM^2*Sqrt2     |LG*( p3.m2*(p1.m3*m1.M2-p1.M2*m1.m3)
                                        |     +m3.m2*(p1.M2*p3.m1-p1.p3*m1.M2))                                                                                                                                                                                                                                                                                                                                                                                                                                                                                                                                                             
vm   |VM   |G.t  |  -GG*Sqrt2/MVM^2     |MVM^2*(1-KG)*m1.m3*m2.M3 
                                        |-LG*( p1.M3*m1.m2*p2.m3-p1.m2*p2.m3*m1.M3
                                        |     -p1.M3*m2.m3*p2.m1+p1.p2*m1.M3*m2.m3)                                                                                                                                                                                                                                                                                                                                                                                                                                                                                                                             
vm   |VM   |G.T  |-i*GG*Sqrt2/MVM^2     |MVM^2*(1-KG)*m1.m3*m2.M3 
                                        |-LG*( p1.M3*m1.m2*p2.m3-p1.m2*p2.m3*m1.M3
                                        |     -p1.M3*m2.m3*p2.m1+p1.p2*m1.M3*m2.m3)                                                                                                                                                                                                                                                                                                                                                                                                                                                                                                                             
vm   |VM.t |G.t  |   2*GG*LG/MVM^2      |m2.m3*(p1.M3*m1.M2-p1.M2*m1.M3)                                                                                                                                                                                                                                                                                                                                                                                                                                                                                                                                                                                              
vm   |VM.t |G.T  | 2*i*GG*LG/MVM^2      |m2.m3*(p1.M3*m1.M2-p1.M2*m1.M3)                                                                                                                                                                                                                                                                                                                                                                                                                                                                                                                                                                                              
vm   |VM.T |G.t  | 2*i*GG*LG/MVM^2      |m2.m3*(p1.M3*m1.M2-p1.M2*m1.M3)                                                                                                                                                                                                                                                                                                                                                                                                                                                                                                                                                                                              
vm   |VM.T |G.T  |  -2*GG*LG/MVM^2      |m2.m3*(p1.M3*m1.M2-p1.M2*m1.M3)                                                                                                                                                                                                                                                                                                                                                                                                                                                                                                                                                                                              
VM   |vm.t |G.t  |   2*GG*LG/MVM^2      |m2.M3*(p1.m3*m1.M2-p1.M2*m1.m3)                                                                                                                                                                                                                                                                                                                                                                                                                                                                                                                                                                                              
VM   |vm.t |G.T  | 2*i*GG*LG/MVM^2      |m2.M3*(p1.m3*m1.M2-p1.M2*m1.m3)                                                                                                                                                                                                                                                                                                                                                                                                                                                                                                                                                                                              
VM   |vm.T |G.t  | 2*i*GG*LG/MVM^2      |m2.M3*(p1.m3*m1.M2-p1.M2*m1.m3)                                                                                                                                                                                                                                                                                                                                                                                                                                                                                                                                                                                              
VM   |vm.T |G.T  |  -2*GG*LG/MVM^2      |m2.M3*(p1.m3*m1.M2-p1.M2*m1.m3)                                                                                                                                                                                                                                                                                                                                                                                                                                                                                                                                                                                              
vm.T |VM.T |G.T  |-i*GG*LG/MVM^2*2*Sqrt2| m1.M3*M1.M2*m2.m3                                                                                                                                                                                                                                                                                                                                                                                                                                                                                                                                                                                                                 
vm.t |VM.T |G.T  |  -GG*LG/MVM^2*2*Sqrt2| m1.M3*M1.M2*m2.m3                                                                                                                                                                                                                                                                                                                                                                                                                                                                                                                                                                                                                 
vm.T | VM.t|G.T  |  -GG*LG/MVM^2*2*Sqrt2| m1.M3*M1.M2*m2.m3                                                                                                                                                                                                                                                                                                                                                                                                                                                                                                                                                                                                                 
vm.t | VM.t|G.T  | i*GG*LG/MVM^2*2*Sqrt2| m1.M3*M1.M2*m2.m3                                                                                                                                                                                                                                                                                                                                                                                                                                                                                                                                                                                                                 
vm.T |VM.T |G.t  |  -GG*LG/MVM^2*2*Sqrt2| m1.M3*M1.M2*m2.m3                                                                                                                                                                                                                                                                                                                                                                                                                                                                                                                                                                                                                 
vm.t |VM.T |G.t  | i*GG*LG/MVM^2*2*Sqrt2| m1.M3*M1.M2*m2.m3                                                                                                                                                                                                                                                                                                                                                                                                                                                                                                                                                                                                                 
vm.T | VM.t|G.t  | i*GG*LG/MVM^2*2*Sqrt2| m1.M3*M1.M2*m2.m3                                                                                                                                                                                                                                                                                                                                                                                                                                                                                                                                                                                                                 
vm.t | VM.t|G.t  |   GG*LG/MVM^2*2*Sqrt2| m1.M3*M1.M2*m2.m3                                                                                                                                                                                                                                                                                                                                                                                                                                                                                                                                                                                                                 
vm.t | VM.t|G    | 2*GG*LG/MVM^2        | M1.M2*(p3.m2*m1.m3-p3.m1*m2.m3)                                                                                                                                                                                                                                                                                                                                                                                                                                                                                                                                                                                               
vm.T | VM.t|G    |2*i*GG*LG/MVM^2       | M1.M2*(p3.m2*m1.m3-p3.m1*m2.m3)                                                                                                                                                                                                                                                                                                                                                                                                                                                                                                                                                                                               
vm.t |VM.T |G    |2*i*GG*LG/MVM^2       | M1.M2*(p3.m2*m1.m3-p3.m1*m2.m3)                                                                                                                                                                                                                                                                                                                                                                                                                                                                                                                                                                                               
vm.T |VM.T |G    | -2*GG*LG/MVM^2       | M1.M2*(p3.m2*m1.m3-p3.m1*m2.m3)                                                                                                                                                                                                                                                                                                                                                                                                                                                                                                                                                                                               
\end{verbatim}
}
\vspace*{-0.2cm}
\caption{\label{tab:lq-inter} An   example of the implementation of
$\LQ$ interactions with gluons in the  CalcHEP/CompHEP
 packages
for the case of $V_{2\mu}$ vector $\LQ$ ($V\!M$) with $F=-2, T_3=-1/2, Q=+1/3$. 
See text for details.}
\vspace*{-0.5cm}
\end{table}
One can see that
the leptoquark Lagrangians
given by Eq.(\ref{eq-lq-glu-scal})--(\ref{eq-lq-glu-vec})
can be rewritten  in terms of products of 
only  three fields. The approach  described above allows us
to represent the strength  tensor as a sum of vector
fields and auxiliary tensor fields. Thus, interactions can be  presented
as a trilinear fields interaction only and, therefore, have simple unambiguously
defined color factor that can be factorized.

As it was demonstrated above,
the implementation of the complete $\LQ$ model 
requires {\it two} auxiliary tensor fields
$\Phi.t$ and $\Phi.T$ for each $\LQ$ type.
In CompHEP~\cite{comp}, only one tensor field ($\Phi.t$) is automatically generated,
while CalcHEP~\cite{calc} generates both of them.
To implement the complete $\LQ$ model into CompHEP
one can introduce the second tensor field
by doubling the vector $\LQ$ fields (by introducing, say, $\tilde\Phi$ $\LQ$ fields), 
and use the tensor field of those particles ($\tilde\Phi.t$)  as a $\Phi.T$ field.
The complete $\LQ$ model given by~Eq.~(\ref{eq:e-q-lq0})-(\ref{eq-lq-glu-vec})
has been  implemented and tested for  both  CompHEP and CalcHEP packages.
The complete models of $\LQ$ interactions relevant to LHC or Tevatron
physics (i.e. models with  $\LQ-l-q$ and  $\LQ-gluon$ interactions)
can be found at \\
\verb|http://hep.pa.msu.edu/belyaev/public/projects/lq/models/lq_calc.zip| \\
and
at \\
\verb|http://hep.pa.msu.edu/belyaev/public/projects/lq/models/lq_comp.zip| \\
for CalcHEP and CompHEP, respectively.
Results for $\LQ$ pair production has been compared 
with those of paper~\cite{Blumlein:1996qp}.
We agree with the results of~\cite{Blumlein:1996qp}
except with the  sign in front of $\lambda_G$
in the expression for vector $\LQ$ pair production (for the sign convention defined by 
Eq.~\ref{eq-lq-glu-vec}),
which we found to be opposite in our calculations.

In  Table~\ref{tab:lq-inter},
we present an explicit  example of the implementation of
$\LQ$ interactions with gluons in the  CalcHEP package
for the case of $V_{2\mu}$ vector $\LQ$ ($V\!M$) with $F=-2, T_3=-1/2, Q=+1/3$ 
(see Table~\ref{tab:lq-table}). Table~\ref{tab:lq-inter}  is a piece of the 
CalcHEP model of particle interactions. 
The first three fields
`$P1$', `$P2$' and  `$P3$'  include the names of the interacting particles.  
The last two fields  `Factor' and  `Lorentz Part' define a
vertex itself. 
Each line of the table represents a particles interaction vertex.
Symbols 'up', 'up.t','up.T' correspond to a vector  $\LQ$, first tensor $\LQ$ and 
second tensor $\LQ$ fields, respectively. 
Names with capital letters: `UP', `UP.t',`UP.T' correspond to the hermitian conjugated
fields. Gluon fields and its two tensor fields are denoted by
`G',`G.t',`G.T', respectively. The `Factor' field
contains `GG', `MUP', `LG' and `Sqrt2' which denote
$g_s$, $M_{UP}$~($\LQ$ mass), $\lambda_G$ and $\sqrt{2}$, respectively.
Symbols `$mi.mj$', `$Mi.mj$', `$mi.Mj$' or `$Mi.Mj$'~$(i,j=1,...3)$
in `Lorentz Part' stand for 
$g^{m_i m_j}$, 
$g^{M_i m_j}$, 
$g^{m_i M_j}$ or  
$g^{M_i M_j}$ metric tensors, respectively, 
with the indices folded with the respective index of vector or tensor particle 
in the column 'i' and 'j'. The first index of the tensor particle
is denoted by a lower case letter, e.g., `$m1$', while the second index
-- by a capital letter, e.g. `$M1$'. The index of the vector particle
is always a lower case letter, e.g. `$m1$'. The symbols `$pi.mj$' or  `$pi.Mj$'
denote `$p_i^{m_j}$' and `$p_i^{M_j}$' i.e. the momenta of the interacting  particles
with the respective Lorentz indices. The symbols `$pi.pj$' denote the
momenta product, $p_i.p_j$, of two particles.
The symbol `KG' in the  `Lorentz Part'
denotes the anomalous coupling $\kappa_G$. A detailed explanation on the 
implementation of models of particle interactions can be found in 
the CompHEP manual~\cite{comp}.

\section{Signal rates}

\FIGURE[h]{
\unitlength=1.5 pt
\SetScale{1.5}
\SetWidth{0.7}      
{} \qquad\allowbreak
\noindent
\begin{picture}(95,79)(0,0)
\Text(15.0,70.0)[r]{$q$}
\ArrowLine(16.0,70.0)(37.0,60.0) 
\Text(15.0,50.0)[r]{$\bar{q}$}
\ArrowLine(37.0,60.0)(16.0,50.0) 
\Text(47.0,67.0)[b]{g}
\Gluon(37.0,60.0)(58.0,60.0){3}{3} 
\Text(80.0,70.0)[l]{$\LQ$}
\DashArrowLine(58.0,60.0)(79.0,70.0){3.0} 
\Text(80.0,50.0)[l]{$\overline{\LQ}$}
\DashArrowLine(79.0,50.0)(58.0,60.0){3.0} 
\end{picture} \
\begin{picture}(95,79)(0,0)
\Text(15.0,70.0)[r]{g}
\Gluon(16.0,70.0)(37.0,60.0){3}{3} 
\Text(15.0,50.0)[r]{g}
\Gluon(16.0,50.0)(37.0,60.0){3}{3} 
\Text(47.0,67.0)[b]{g}
\Gluon(37.0,60.0)(58.0,60.0){3}{3} 
\Text(80.0,70.0)[l]{$\LQ$}
\DashArrowLine(58.0,60.0)(79.0,70.0){3.0} 
\Text(80.0,50.0)[l]{$\overline{\LQ}$}
\DashArrowLine(79.0,50.0)(58.0,60.0){3.0} 
\end{picture} \ 
\\
\begin{picture}(95,79)(0,0)
\Gluon(10.0,65.0)(40.0,65.0){3.0}{3}      \Text(5.0,70.0)[r]{g}
\DashArrowLine(40.0,65.0)(70.0,65.0){3.0} \Text(66.0,70.0)[r]{$\LQ$}
\DashArrowLine(40.0,65.0)(40.0,45){3.0}   \Text(65.0,50.0)[l]{$\overline{\LQ}$}
\DashArrowLine(40.0,45.0)(68.0,45.0){3.0} \Text(45,55.0)[l]{$\LQ$}
\Gluon(40.0,45.0)(10.0,45.0){3.0}{3}      \Text(5.0,45.0)[r]{g}
\end{picture} \
\begin{picture}(95,79)(0,0)
\Gluon(10.0,70)(30.0,60.0){3.0}{3}        \Text(5.0,70.0)[r]{g}
\Gluon(30.0,60)(10,45){3.0}{3} 	   	  \Text(5.0,50.0)[r]{g}
\DashArrowLine(30.0,60.0)(60.0,70.0){3.0} \Text(65.0,70.0)[l]{$\LQ$}
\DashArrowLine(60.0,45.0)(30.0,60.0){3.0} \Text(65.0,45.0)[l]{$\overline{\LQ}$}
\end{picture}
\caption{\label{diag-lq-pair}
Diagrams for $\LQ$ pair production in quark-quark and gluon-gluon fusion.}
}

\FIGURE[h]{
\unitlength=1.5 pt
\SetScale{1.5}
\SetWidth{0.7}      
{} \qquad\allowbreak
\noindent
\begin{picture}(95,79)(0,0)
\Text(15.0,70.0)[r]{$q$}
\ArrowLine(16.0,70.0)(37.0,60.0) 
\Text(15.0,50.0)[r]{g}
\Gluon(37.0,60.0)(16.0,50.0){3}{3} 
\Text(47.0,67.0)[b]{q}
\ArrowLine(37.0,60.0)(58.0,60.0)
\Text(80.0,70.0)[l]{$\bar{\ell}$}
\ArrowLine(58.0,60.0)(79.0,70.0)
\Text(80.0,50.0)[l]{$\LQ$}
\DashArrowLine(79.0,50.0)(58.0,60.0){3.0} 
\end{picture} \
\begin{picture}(95,79)(0,0)
\Gluon(10.0,65.0)(40.0,65.0){3.0}{3}      \Text(5.0,70.0)[r]{g}
\DashArrowLine(40.0,65.0)(70.0,65.0){3.0} \Text(66.0,70.0)[r]{$\LQ$}
\DashArrowLine(40.0,65.0)(40.0,45){3.0}   \Text(45.0,55.0)[l]{$\overline{\LQ}$}
\ArrowLine(40.0,45.0)(70.0,45.0)          \Text(68,40.0)[l]{$\bar{\ell}$}
\ArrowLine(40.0,45.0)(3,45.0)             \Text(0,45.0)[r]{q}
\end{picture} 
\caption{\label{diag-lq-single}
Diagrams for $\LQ$ single production in quark-gluon fusion}
}

Leptoquarks can be produced in quark-quark, quark-gluon
and gluon-gluon interactions. Their production at the LHC is dominated by
the strong interaction of $\LQ$ with gluons, and 
one can safely neglect $\LQ$ interactions with photons, Z and $W^\pm$ bosons.
In the case of  quark-quark and gluon-gluon fusion, $\LQ$ produced {\it in pairs}
are shown by Feynman diagrams in Fig.~\ref{diag-lq-pair}.
The $\LQ$ pair production cross section is defined by $\LQ$ mass and 
coupling of the strong interactions -- $\alpha_s$.
In the case of   {\it vector} $\LQ$ pair production the total cross section
also depends on the anomalous $\kappa_G$ and $\lambda_G$ couplings.

In the case of quark-gluon fusion,
$\LQ$ are produced {\it singly} in association with leptons.
In comparison  to $\LQ$ pair production, the production rate of 
$\LQ$ single rate
depends also   on
$\LQ-quark-lepton$ Yukawa type coupling appearing in 
 Eq.~(\ref{eq:e-q-lq0})-(\ref{eq:e-q-lq2}).
The corresponding diagrams are shown in Fig.~\ref{diag-lq-single}.

Both processes --- $\LQ$ single and pair production ---
give the same striking signatures:\\
1. $2l+jets$ when  $\LQ$s  decay into a lepton and a quark;\\
2. $l+jets+\emis$ for both processes when singly produced $\LQ$ decays into a neutrino and a quark
and  one of the $\LQ$ produced in pair production, decays into a lepton and a quark
while the other decays into a neutrino and a quark.
The same signature will take place for $\LQ$ single production 
if $\LQ$  is produced in association with a neutrino
but decays into a lepton and a quark.
\\
3. $jets+\emis$ signature when  $\LQ$s decay into a neutrino and a quark
and single $\LQ$ is produced in association with a neutrino.

One of the important message which we would like to convey in this article is 
that $\LQ$ single and pair productions must be considered together.
Because of similarity in their signatures,
it is impossible to  separate  signals from $\LQ$ single and pair production completely.

In our study, for  numerical calculations  we have used the CTEQ6L set for parton 
distribution function (PDF)\cite{cteq6} and chosen the QCD scale
equal to the $\LQ$ mass. 
There are many scalar and vector $\LQ$ species with different isospin, charge
and  $\LQ-quark-lepton$ couplings
(as given in Table~\ref{tab:lq-table}), however,
the total $\LQ$ decay width depends on
one parameter~(given that $\LQ$ mass and $\alpha_S$ are fixed),
the sum of $\LQ-quark-lepton$ couplings squared,
$\lambda_{eff}^2 = \lambda_L^2({lq})
+\lambda_R^2({lq})+\lambda_L^2({\nu q'})$~(see Table~\ref{tab:lq-table} 
and Eq.~(\ref{eq:e-q-lq0}),~(\ref{eq:e-q-lq2}).
Without loosing generality, for our analysis, we have chosen  scalar($S0$) and  vector($V\!M$) $\LQ$
(see Table~\ref{tab:lq-table}), which couples to  both -- the charged lepton and a quark 
as well as to a neutrino and a quark. 
For the reference point, we have chosen 
$\lambda_{eff}$ to be 
equal to the electromagnetic coupling, $e=\sqrt{4\pi\alpha}\simeq 0.312$  for  $\alpha=1/128$.
Choosing $\lambda_{eff}$ completely defines $\LQ$ decay width:
 \begin{equation}
\mbox{\ \ \ \ \ \ \ \ \ $S0$ with} \ \ F=-2, T_3=\ \ \ \ 0, \ \ Q=+1/3 
\mbox{\ has }\ \Gamma= M_\LQ {(\lambda_L^2(lq)+\lambda_L^2(\nu q')) \over16\pi},
\end{equation}
\begin{equation}
\mbox{while $V\!M$ with} \  \ F=-2, T_3=-1/2, \ \ Q=+1/3 
\mbox{\ has }\ \Gamma= M_\LQ {(\lambda_R^2(lq)+\lambda_L^2(\nu q') \over 24 \pi}.
\end{equation}
Both, $S0$ and $V\!M$, decay into  $\bar{u} e^+$  and $\bar{d}\bar{\nu}$
with 50\% branching fraction for each decay channel for our choice of parameters.
For $M_{\LQ}=1$~TeV, $\sqrt{g_{3L}^2+g_{3L}^2}=\sqrt{g_{2R}^2+g_{2L}^2}=\lambda_{eff}=e$,
the total width for $S0$ is $\Gamma(S0)=1.95$~GeV
while $\Gamma(VM)=1.30$~GeV. 
 
\FIGURE{
\includegraphics[width=7.8cm,height=7.8cm]{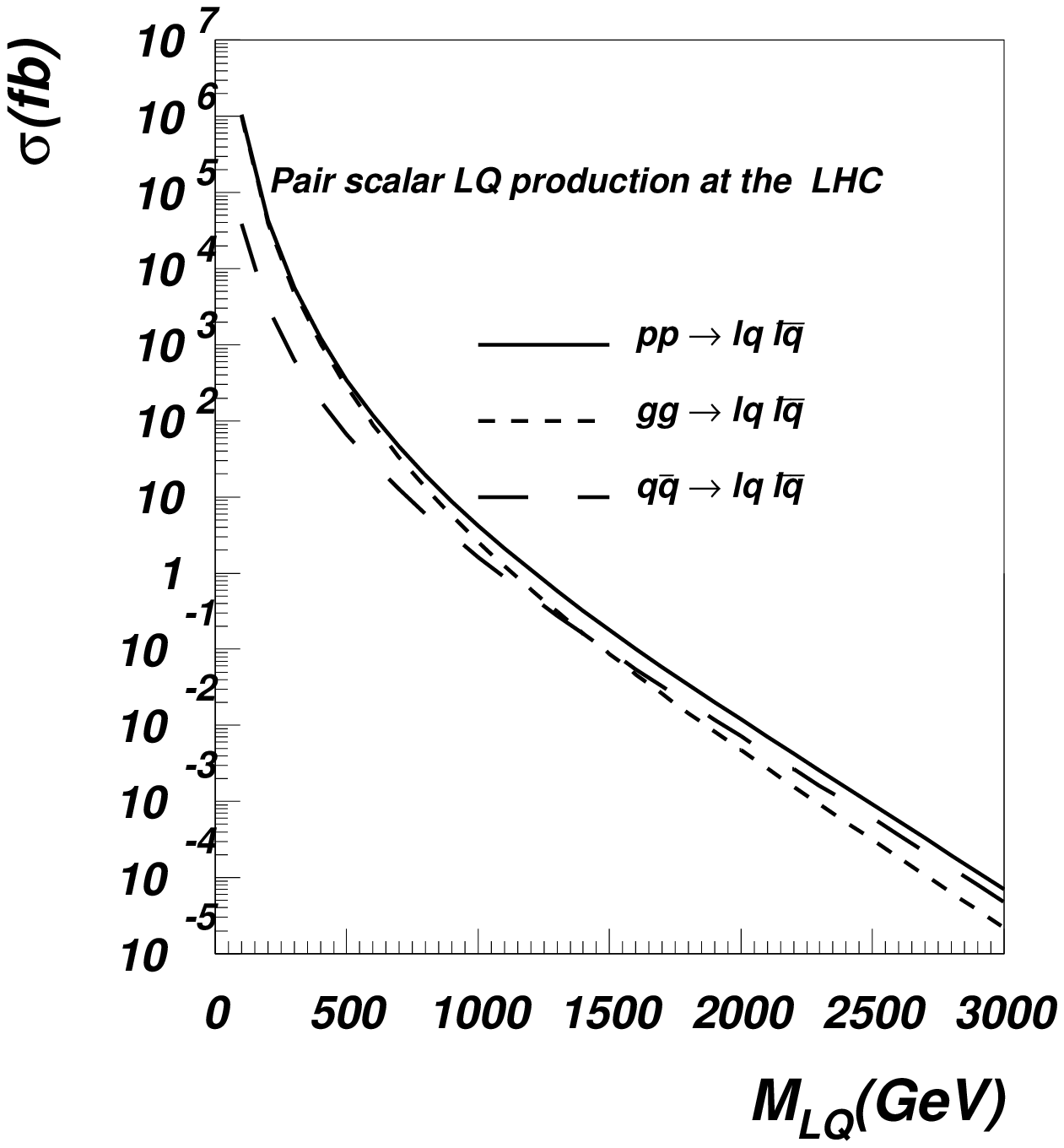}%
\includegraphics[width=7.8cm,height=7.8cm]{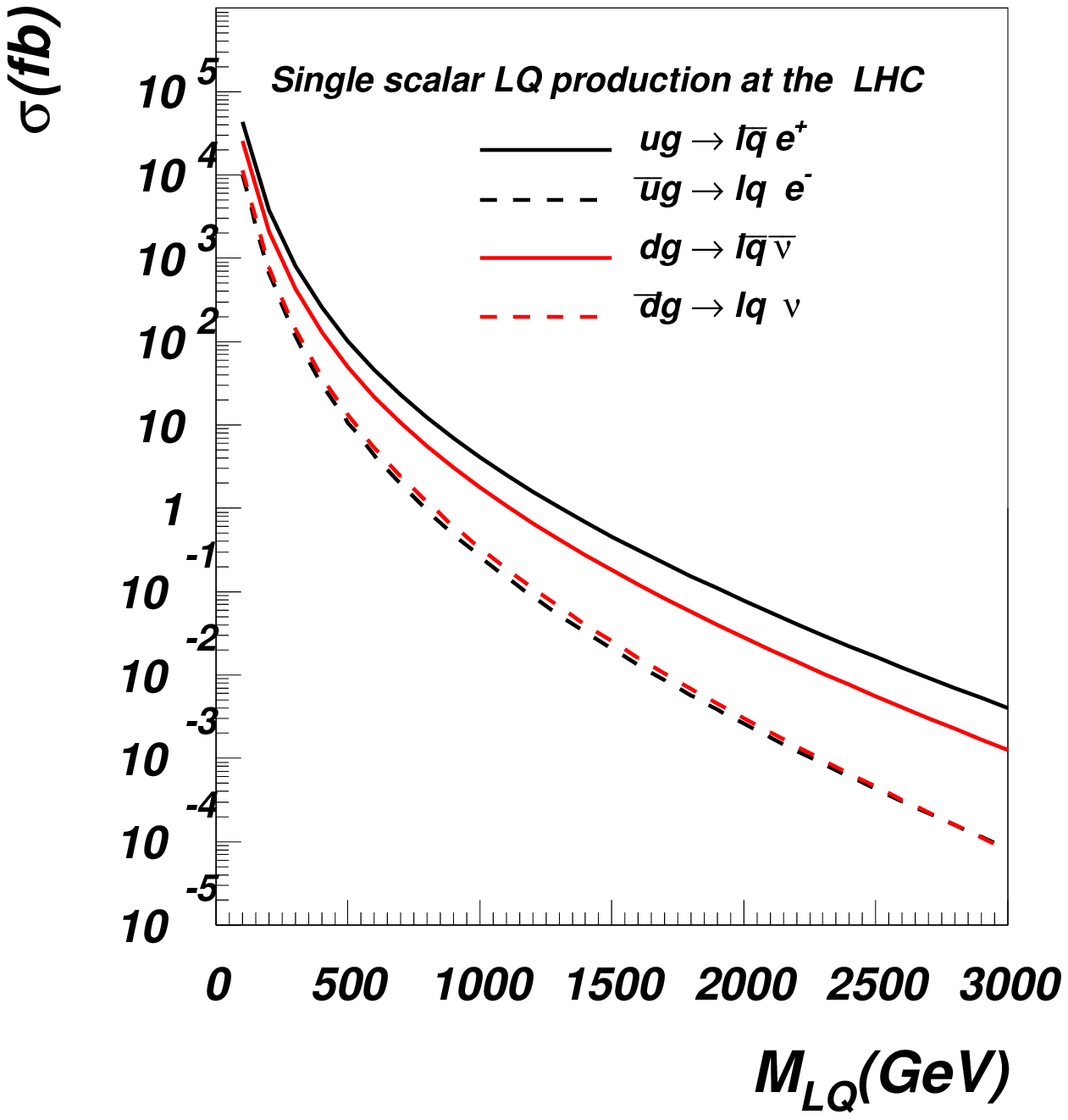}
\caption{\label{fig:cs-slq}
Total cross section of scalar $\LQ$ ($S_0$) production as a function of the $\LQ$ mass
for $\LQ$ pair~(left) and single~(right) production 
at the LHC.
The cross section of single $\LQ$  production depends on the value
$\LQ-\ell-q$ coupling chosen to be $\sqrt{g_{3L}^2+g_{3L}^2}=\lambda_{eff}=e$.
The single $\LQ$
production  in association  with charged lepton is represented by black
curves  while its production  in association with  neutrino
is shown by red curves.
} 
} 

In Fig.~\ref{fig:cs-slq}
we present the
total cross section of  scalar $\LQ$ production as a function of $\LQ$ mass
for $\LQ$ pair~(left) and single~(right) production.
One should note, that
single  $\LQ$ and single $\overline\LQ$ production rates
at the LHC are drastically different
since the first of them  is initiated by valence quarks while the other one -- by 
sea quarks.
Below, we quote the sum of the cross sections for $\LQ$ and $\overline\LQ$ single 
production.
$\LQ$ single production rates also depend on to which quark(s)
it couples. In Fig.~\ref{fig:cs-slq} one can  see the difference in rates for single  $\LQ$
production when it is produced in association  with charged lepton (higher rates) 
and when it is produced in association with a neutrino (lower rates).
The apparent origin of this difference lays in the 
difference between $up-$ and $down-$ quark parton densities,
respectively. 
We do not lose generality when studying  just one type of scalar $\LQ$,  $S0$ here,
since the generic cross section of the scalar $\LQ$ single  production could be expressed
as a superposition of two cross sections represented by the red and black curves in 
Fig.~\ref{fig:cs-slq} which scales quadratically with the $\LQ-l-q$ coupling.
The same is also true for the case of vector $\LQ$ production
which we describe below.

For low $\LQ$ masses ($\sim 100$~GeV) $\LQ$ pair production 
process dominates over the $\LQ$ single production process by
about one order of magnitude: $\sim 1000$~pb compared with $\sim 100$~pb.
But for larger $\LQ$ masses ($\sim 1000$~GeV), targeted by LHC (unless Tevatron will 
discover light $\LQ$ states earlier)
the cross sections of $\LQ$ single and pair production become
comparable. 
This happens because of a stronger phase space suppression of $\LQ$ pair production
and a related faster drop of the  parton (especially gluon)
density functions. For $M_{\LQ}\sim 1000$ GeV,
the cross section for both, single  and $\LQ$ pair production, is roughly
of the order of 
$10$~fb. For even higher scalar $\LQ$ masses, single production
starts to dominate over $\LQ$ pair production --
for $M_{\LQ}\sim 2000$ GeV, the $\LQ$ single production cross section ($\sim 0.1$~fb) 
is already one order of magnitude larger than the cross section of pair $\LQ$
production ($\sim 0.01$~fb)! Of course, the cross section of $\LQ$ single production
is proportional to the square of the $\LQ-e-q$ unknown coupling 
(chosen to be equal to the electromagnetic coupling)
and can be rescaled accordingly.
One can see that if this coupling is of the order of the electromagnetic coupling
(like in the case of recent experimental limits obtained at HERA~\cite{lq-hera}), 
then the scalar $\LQ$ single  production should be definitely taken into account
and combined with the studies of the $\LQ$ pair production.
\FIGURE{
\includegraphics[width=7.8cm,height=7.8cm]{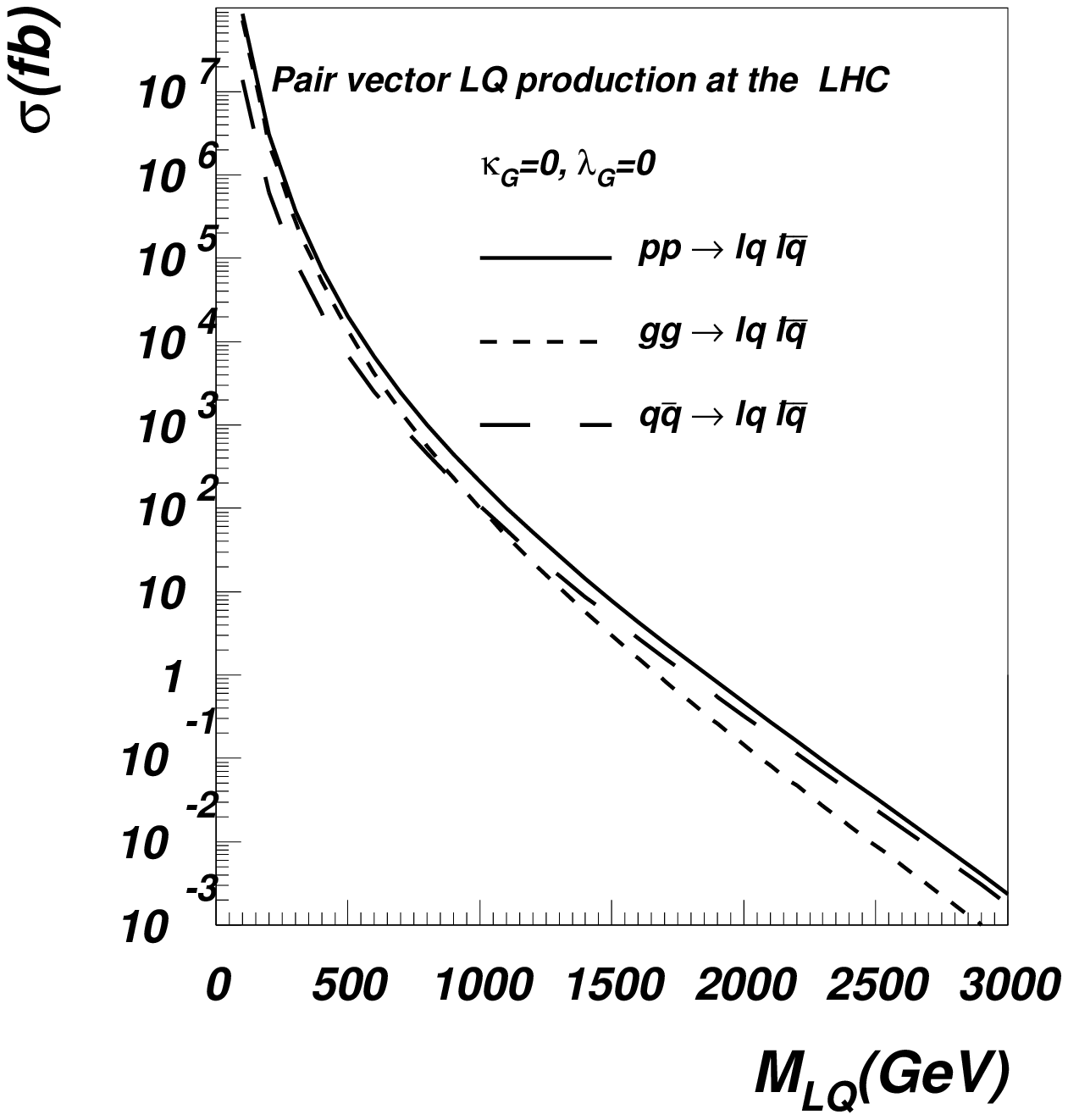}%
\includegraphics[width=7.8cm,height=7.8cm]{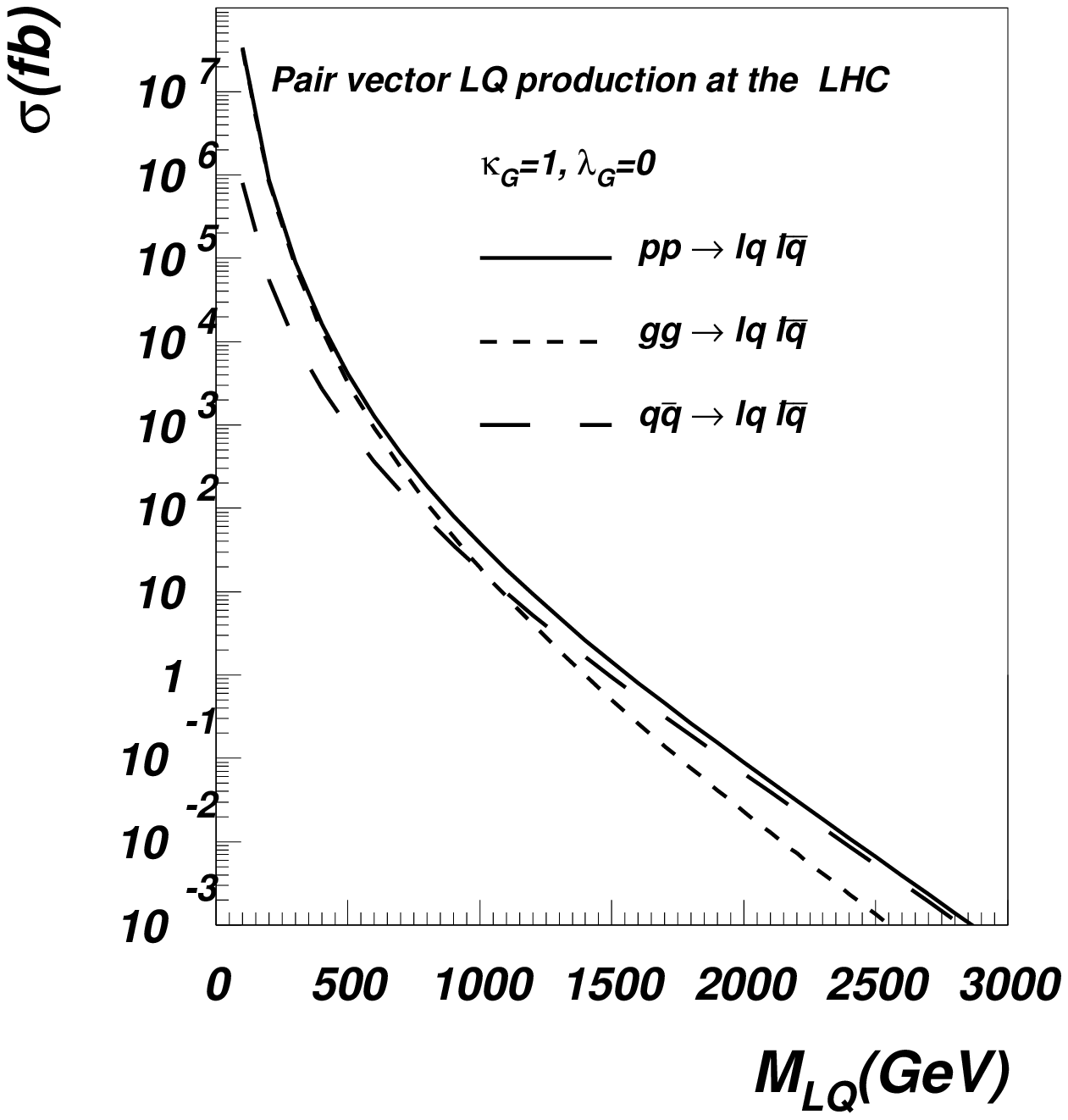}
\includegraphics[width=7.8cm,height=7.8cm]{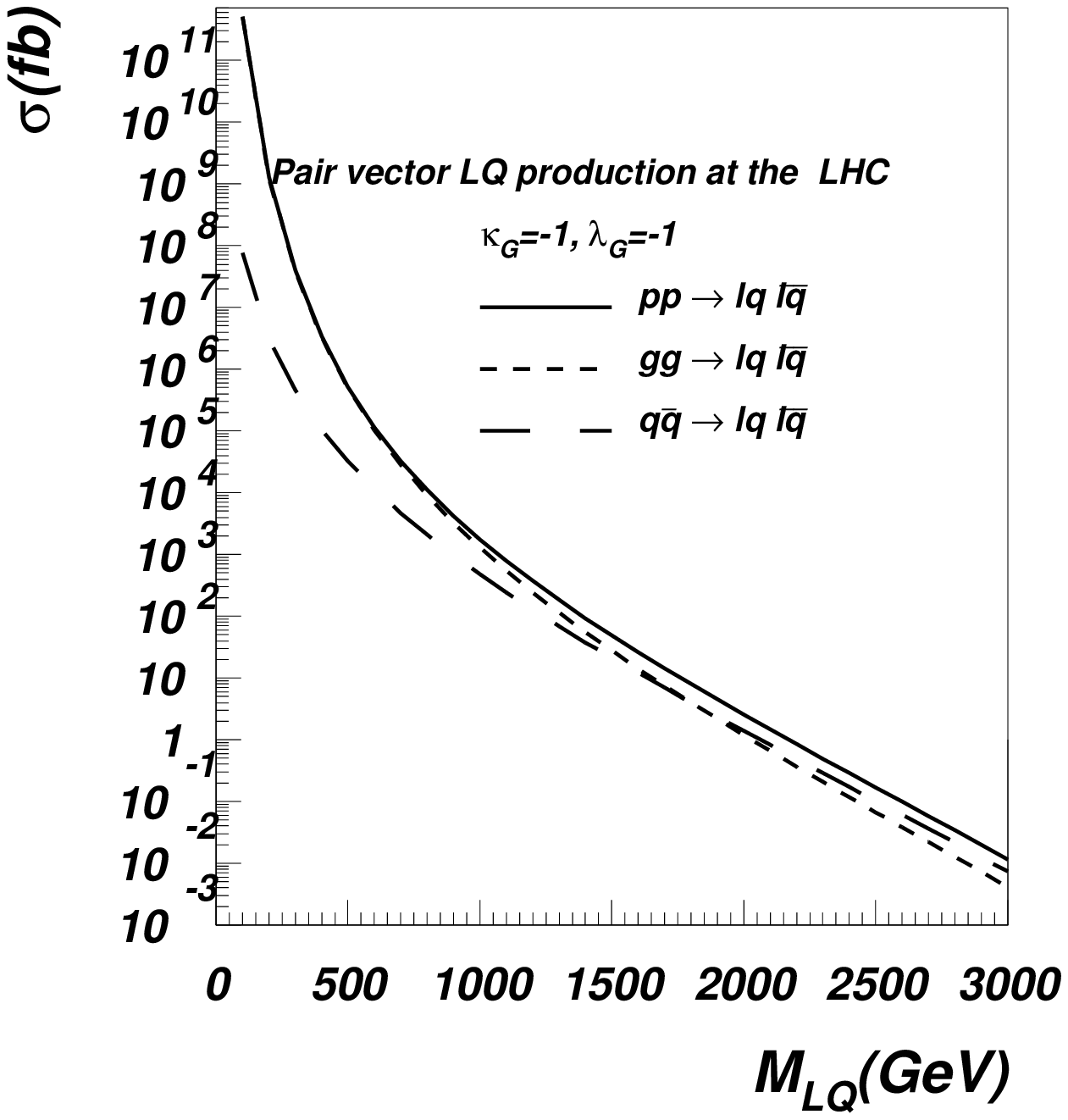}%
\includegraphics[width=7.8cm,height=7.8cm]{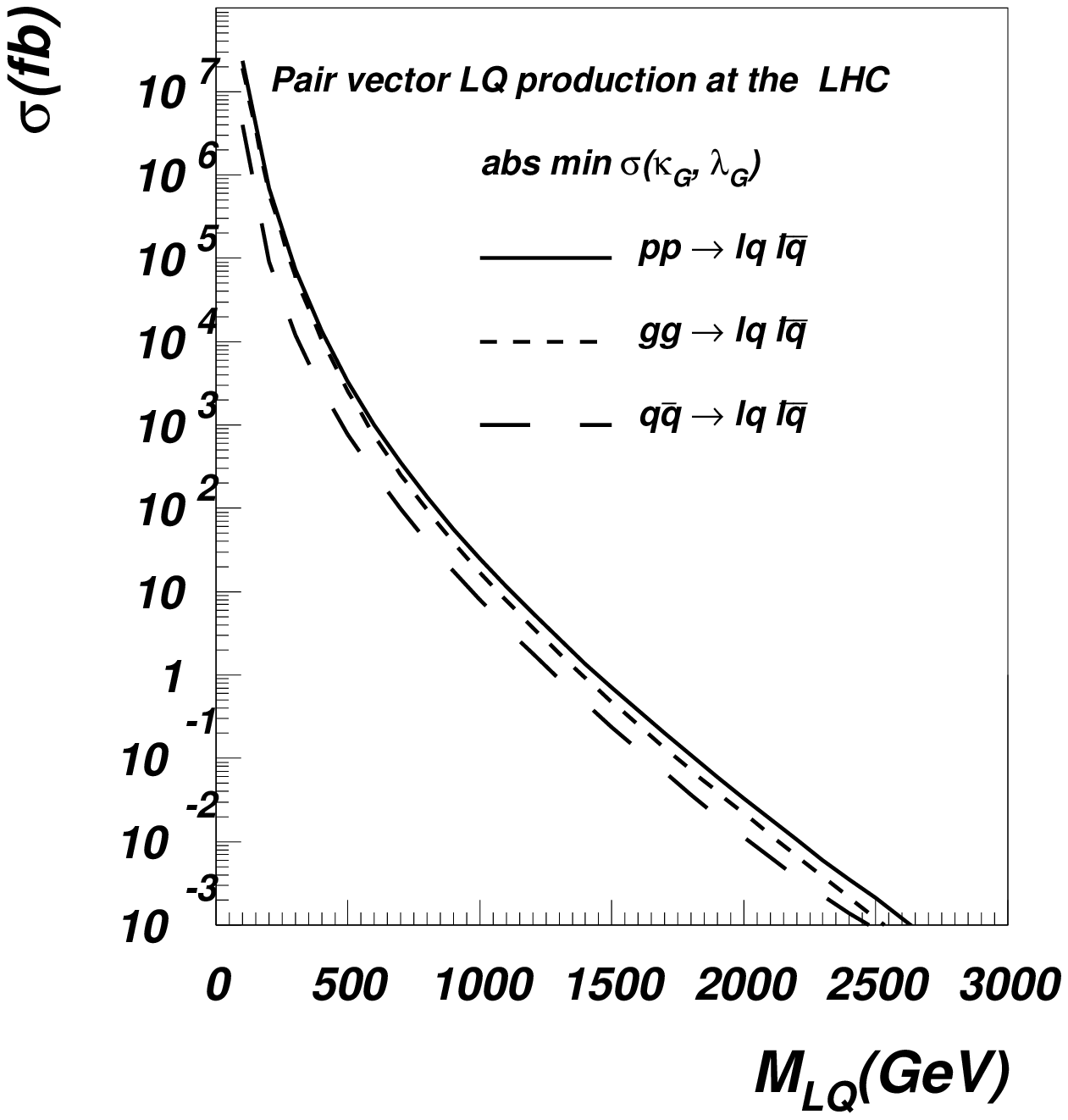}
\caption{\label{fig:cs-vlq-pair}
Total cross section for pair vector $\LQ$ production as a function of the $\LQ$ mass.
Four choices of ($\kappa,\lambda$) are presented:
1) $\kappa_G = \lambda_G = 0$   -- Yang-Mills type coupling (YM);
2) $\kappa_G=1$, $\lambda_G=0$ -- Minimal coupling (MC);
3) $\kappa_G=-1$, $\lambda_G=-1$ -- (MM) and
4) the case of {\it absolute minimal cross section} (AM)
   in which the cross-section is minimized with respect to  $\kappa_G,\lambda_G$
   parameters for each value of  $M_\LQ$.}  
} 
 
\FIGURE{
\includegraphics[width=7.8cm,height=7.8cm]{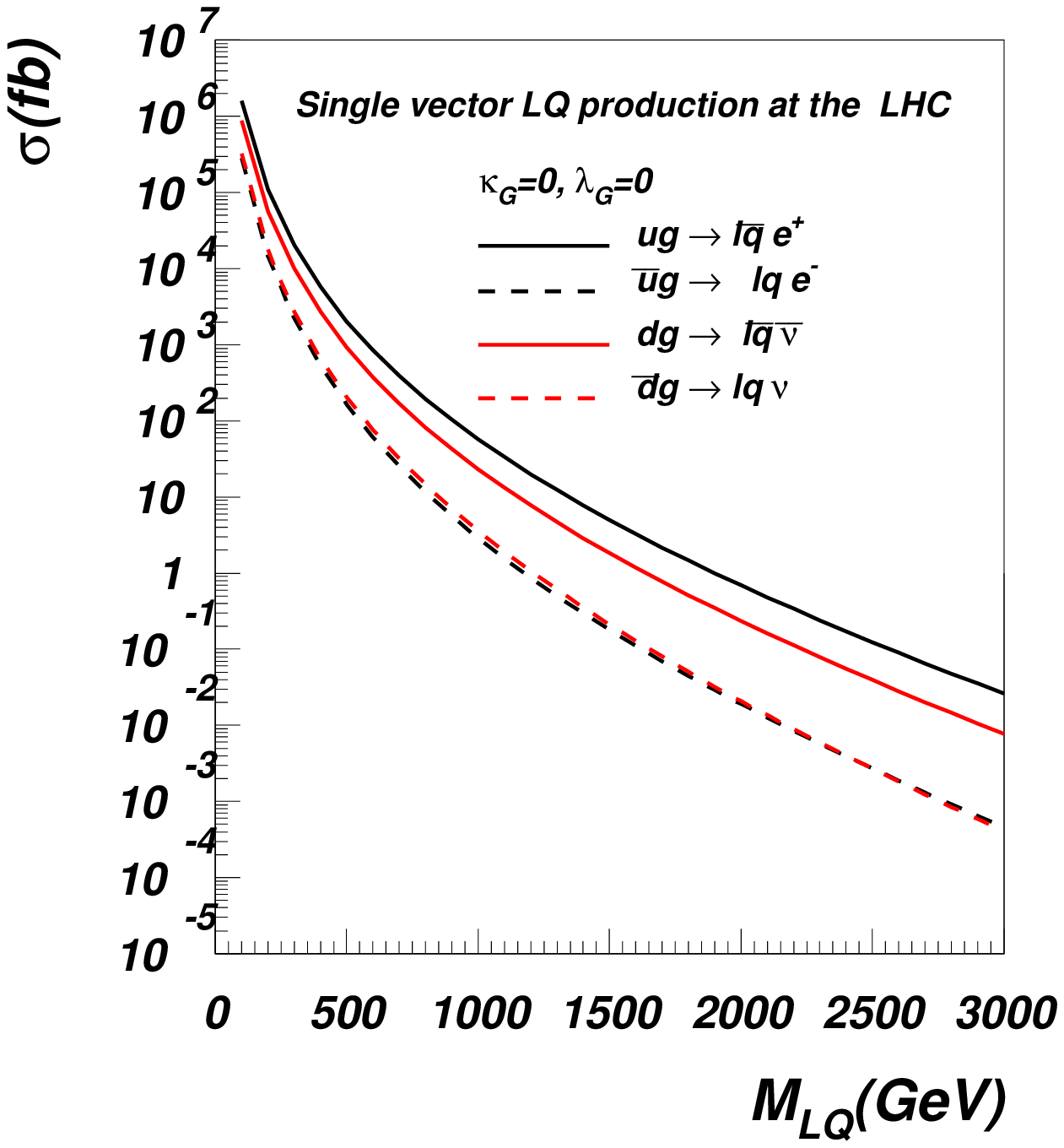}%
\includegraphics[width=7.8cm,height=7.8cm]{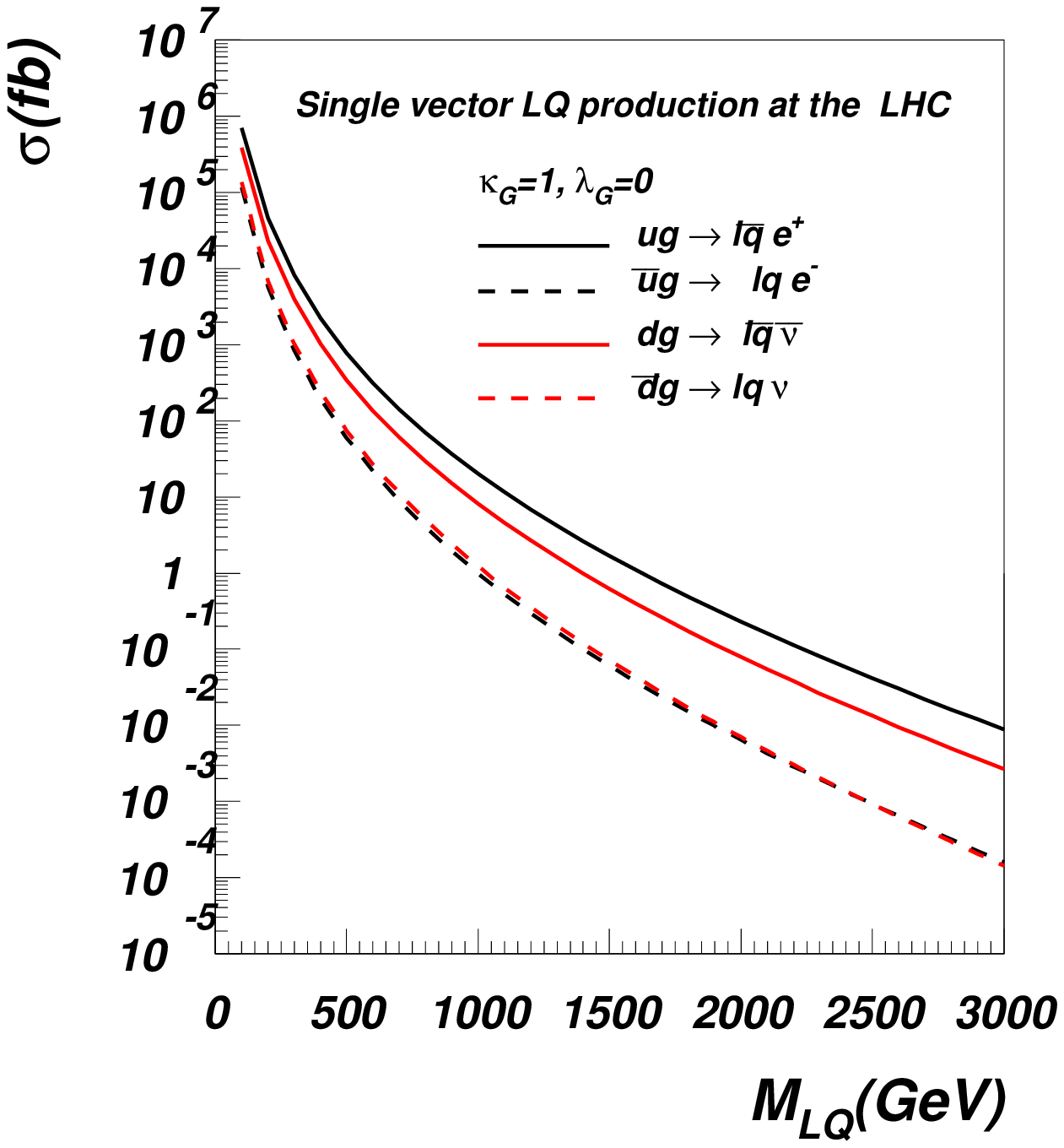}
\includegraphics[width=7.8cm,height=7.8cm]{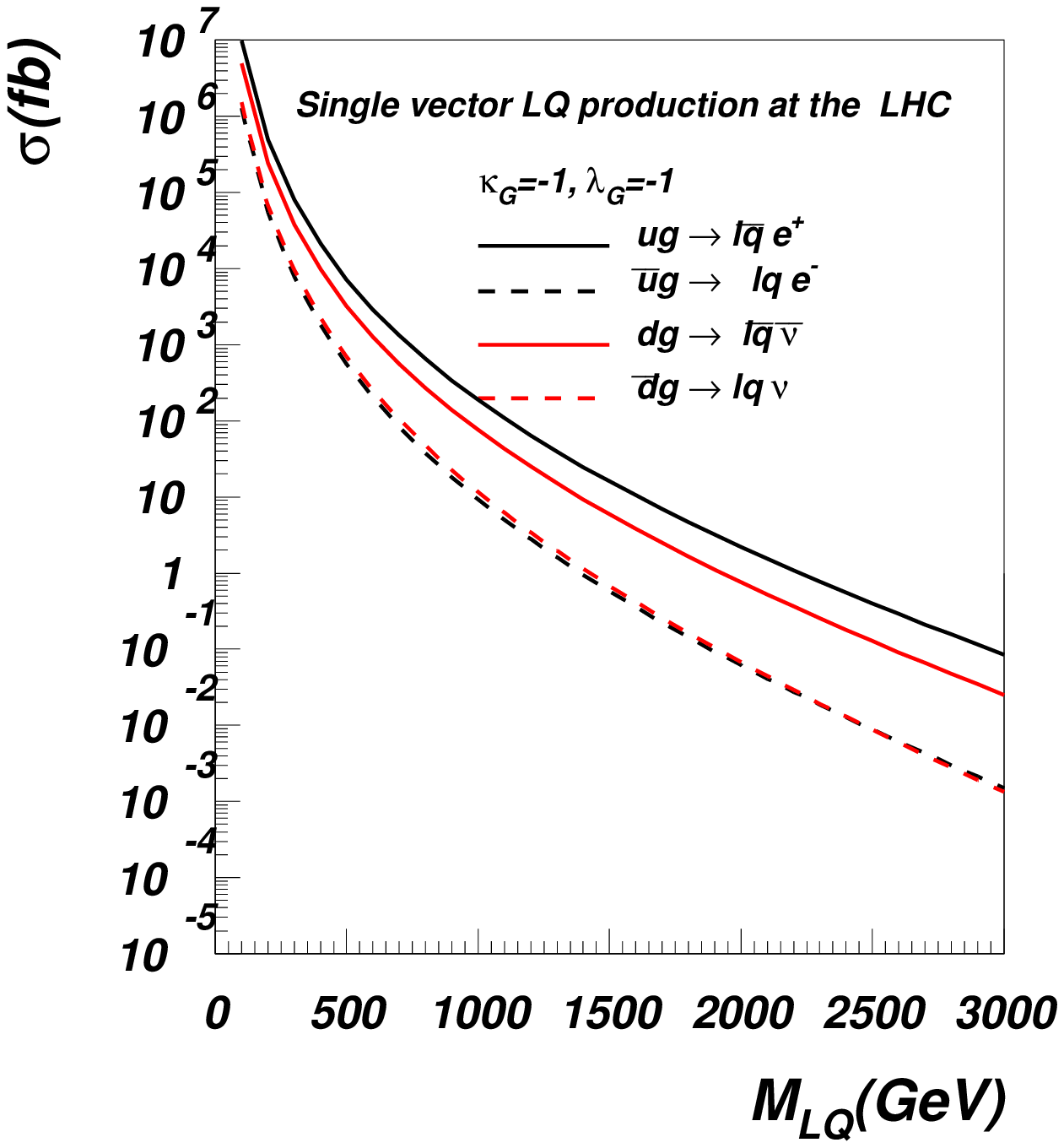}%
\includegraphics[width=7.8cm,height=7.8cm]{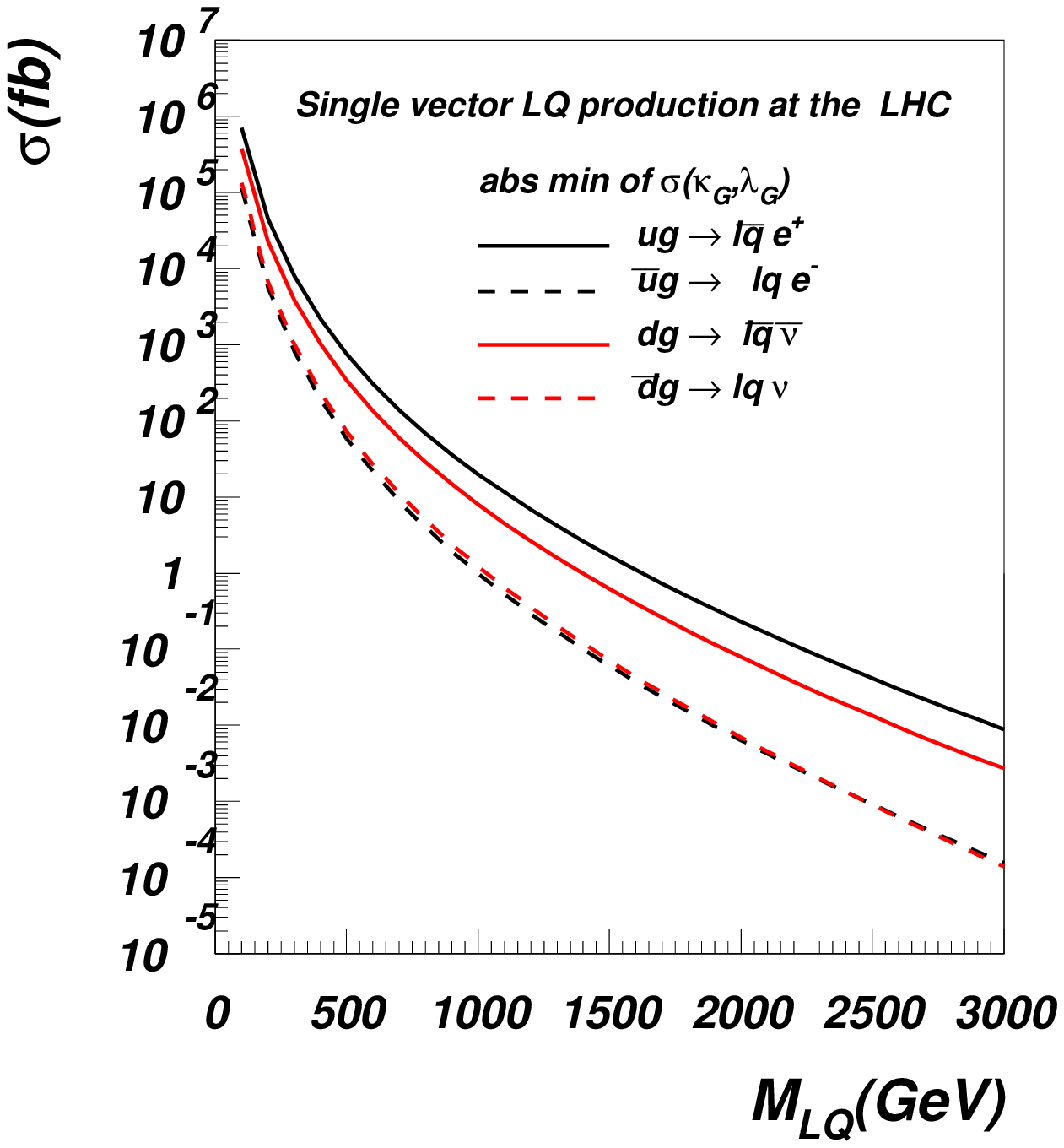}
\caption{\label{fig:cs-vlq-sngl}
Total cross section of single vector $\LQ$($V\!M$)  production as a function of the $\LQ$ mass.
The cross section of $\LQ$ single production depends on its quark content and the value
$\LQ-\ell-q$ coupling chosen to be $\sqrt{g_{2R}^2+g_{2L}^2}=\lambda_{eff}=e$.
single $\LQ$
production  in association  with the charged lepton is denoted by black
curves, while its production  in association with neutrino
is denoted by red curves.
Four choices of ($\kappa,\lambda$) are presented:
1) $\kappa_G = \lambda_G = 0$   -- Yang-Mills type coupling (YM);
2) $\kappa_G=1$, $\lambda_G=0$ -- Minimal coupling (MC);
3) $\kappa_G=-1$, $\lambda_G=-1$ -- (MM) case and
4) the case of {\it absolute minimal cross section} (AM)
   in which the cross section is minimized with respect to  $\kappa_G,\lambda_G$
   parameters for each value of  $M_\LQ$.}} 

Vector $\LQ$ cross sections are presented 
in Fig.~\ref{fig:cs-vlq-pair} and Fig.~\ref{fig:cs-vlq-sngl}
for the cases of $\LQ$ single and pair production, respectively.
We present results for four `traditional' choices of $\kappa_G$ and $\lambda_G$:
\\
1. $\kappa_G = \lambda_G = 0$, Yang-Mills type coupling (YM) case \\
2. $\kappa_G=1$, $\lambda_G=0$, Minimal coupling (MC) case\\
3. $\kappa_G=-1$, $\lambda_G=-1$, MM case\\
4. the case of {\it absolute minimal cross section} (AM)
   in which the cross section is minimized with respect to  $\kappa_G,\lambda_G$
   parameters for each value of  $M_\LQ$.
\\
For AM case,  one can establish the absolute conservative limit
on the cross section of vector $\LQ$ 
since for each value of $M_\LQ$ the minimal cross section is defined.
One can see again,
that the cross section for $\LQ$ single production catches up the cross section of 
pair production at  $M_\LQ\simeq 1$~TeV and starts to dominate at larger $\LQ$ masses
for the same reason as in scalar $\LQ$ production case.
One can also notice that the cross section for MC and AM cases
are quite close to each other for $\LQ$ masses $\leq$~1 TeV, e.g.at  $M_\LQ =  1.0$~TeV
they are 37.6~fb (MC) and 24.8f~b (AM)   for pair vector $\LQ$
production and 31.4~fb (MC) and 30.2~fb (AM) for 
vector $\LQ$ single production. Here, we present the sum of the cross sections, $pp\to \LQ\ell$ and 
$pp\to \LQ\nu$,  of  $\LQ$ single production processes.
However, at higher $\LQ$ masses~(1.5-2~TeV) the cross section for $\LQ$ pair production 
in the AM case could be a factor of 2-3 smaller than for MC case. This happens because 
the contribution from $gg\to \LQ\overline{\LQ}$ to the total production cross section
(which is similar for MC and AM cases) vanishes for heavy $\LQ$ production,
while the contribution from $q\bar{q}\to \LQ\overline{\LQ}$ starts  to
dominate (it is several times larger for MC, as compared to AM case)
in this mass region.
\TABLE{
\begin{tabular}{|c|c|c|c|c|cc|}
\hline\hline
&\multicolumn{6}{|c|}{$\LQ$ pair production}\\
\cline{2-7}
$M_\LQ$ & Scalar & \multicolumn{5}{|c|}{Vector ($\kappa,\lambda)$}\\
\cline{3-7}
 (TeV)  &        & MM	    & YM      &   MC   & AM      & ($\kappa,\lambda$)\\
        &        & $(-1,-1)$& $(0,0)$ & $(1,0)$&         &  		    \\
\hline
  0.1    &1.04E+06&5.05E+11  &8.55E+07&3.34E+07   &2.31E+07 & (0.549, 0.00363)\\
  0.5    &3.45E+02&5.23E+05  &2.02E+04&4.12E+03   &3.31E+03 & (1.01,  0.0496)\\
  1.0    &4.20E+00&1.76E+03  &2.08E+02&3.76E+01   &2.48E+01 & (1.16,  0.130)\\
  1.5    &1.79E-01&4.89E+01  &7.84E+00&1.44E+00   &7.12E-01 & (1.25,  0.198)\\
  2.0    &1.19E-02&2.57E+00  &4.70E-01&8.96E-02   &3.32E-02 & (1.30,  0.239)\\
\hline
\hline
& \multicolumn{6}{|c|}{$\LQ$ single production: $pp(ug)\to \LQ \ell$}\\
\cline{2-7}
$M_\LQ$ & Scalar & \multicolumn{5}{|c|}{Vector ($\kappa,\lambda)$}\\
\cline{3-7}
 (TeV)  &        & MM	    & YM      &   MC   & AM      & ($\kappa,\lambda$)\\
        &        & $(-1,-1)$& $(0,0)$ & $(1,0)$&         &		     \\
\hline
  0.1   &5.35E+04&1.12E+07  &1.87E+06 &8.25E+05&8.15E+05 & (0.991, 0.0381)   \\
  0.5   &1.12E+02&7.80E+03  &2.22E+03 &8.35E+02&8.20E+02 & (1.06,  0.0792)   \\
  1.0   &4.37E+00&1.97E+02  &6.05E+01 &2.21E+01&2.10E+01 & (1.06,  0.0656)   \\
  1.5   &4.79E-01&1.66E+01  &5.15E+00 &1.75E+00&1.74E+00 & (1.05,  0.0215)   \\
  2.0   &1.60E-01&2.29E+00  &7.10E-01 &2.38E-01&2.37E-01 & (1.04, -0.0437)   \\
\hline
\hline
& \multicolumn{6}{|c|}{$\LQ$ single production: $pp(dg)\to \LQ \nu$}\\
\cline{2-7}
$M_\LQ$ & Scalar & \multicolumn{5}{|c|}{Vector ($\kappa,\lambda)$}\\
\cline{3-7}
 (TeV)  &        & MM	    & YM      &   MC   & AM      & ($\kappa,\lambda$)\\
        &        & $(-1,-1)$& $(0,0)$ & $(1,0)$&         &		     \\
\hline
  0.1   &3.71E+04&6.55E+06  &1.21E+06 &5.20E+05&5.15E+05 & (1.01,  0.0431)   \\
  0.5   &6.30E+01&3.93E+03  &1.14E+03 &4.23E+02&4.15E+02 & (1.06,  0.0790)   \\
  1.0   &2.11E+00&8.70E+01  &2.68E+01 &9.30E+00&9.20E+00 & (1.06,  0.0554)   \\
  1.5   &2.06E-01&6.60E+00  &2.05E+00 &6.90E-01&6.90E-01 & (1.05,  0.0429)   \\
  2.0   &3.13E-02&8.25E-01  &2.56E-01 &8.60E-02&8.55E-02 & (1.04, -0.0828)   \\
\hline
\hline
\end{tabular}
\caption{\label{tab:lq-cs}
 Cross sections (in fb)
for $\LQ$ single and pair production 
for  vector ($V\!M$) and scalar ($S0$) leptoquarks at the LHC.
The cross section of $\LQ$ single production depends on its quark content and the value
of $\LQ-\ell-q$ coupling chosen to be $\sqrt{g_{2R}^2+g_{2L}^2}=\lambda_{eff}=e$.
single $\LQ$
production in association  with a neutrino is presented separately from 
$\LQ$ production  in association with a charged lepton.
Four choices for ($\kappa,\lambda$) are presented for  vector $\LQ$ production:
1) $\kappa_G = \lambda_G = 0$   -- Yang-Mills type coupling (YM);
2) $\kappa_G=1$, $\lambda_G=0$ -- Minimal coupling (MC);
3) $\kappa_G=-1$, $\lambda_G=-1$ -- (MM) case and
4) the case of {\it absolute minimal cross section} (AM)
   in which the cross section is minimized with respect to  $\kappa_G,\lambda_G$
   parameters for each value of  $M_\LQ$.} 
}	       

The cross section for YM case is typically a factor of 3-5 higher 
than the one for MC case.
For example, for $M_\LQ =  1.0$~TeV one has 
208~fb and 87.3~fb for $\LQ$  single and pair 
production cross section for YM case.
 For MM case, the cross section for pair $\LQ$s is 1-3 orders 
of magnitude higher than the one for
MC case: $\sigma_\LQ^{pair}(MM)=1760$~fb  for  $M_\LQ =  1.0$~TeV.
The MM cross section for $\LQ$ single production is about one order 
of magnitude higher, as compared to MC case: : $\sigma_\LQ^{single}(MM)=284$~fb  
for  $M_\LQ =  1.0$~TeV.

In Table~\ref{tab:lq-cs}, we summarize cross sections 
for $\LQ$ single and pair production 
for both, vector ($V\!M$) and scalar ($S0$) leptoquarks at the LHC.
As we mentioned above, the total cross section for any $\LQ$ type 
can be obtained from Table~\ref{tab:lq-cs}
by superimposing respective numbers and rescaling them
for a chosen value of $\LQ-\ell-q$ coupling.


In our study we use the MC set of cross sections (which is not very different from AM 
case) to establish a conservative limit on the $M_\LQ$ and $\lambda_{eff}$ parameters 
of the model.

\section{Simulations and signal versus background analysis}

The simulations of leptoquark signal events were performed with 
CompHEP~\cite{comp}, the CompHEP-PYTHIA interface~\cite{Compyth} and 
PYTHIA6.2~\cite{Pyth} program chain.
The cross section values presented in this article were calculated using CTEQ6L parton 
distribution function (PDF)\cite{cteq6}.
PYTHIA  was used to account for initial and final state radiation and to perform 
hadronization and decay of resonances, when it was relevant.

\FIGURE{
\includegraphics[width=8.5cm,height=8.5cm]{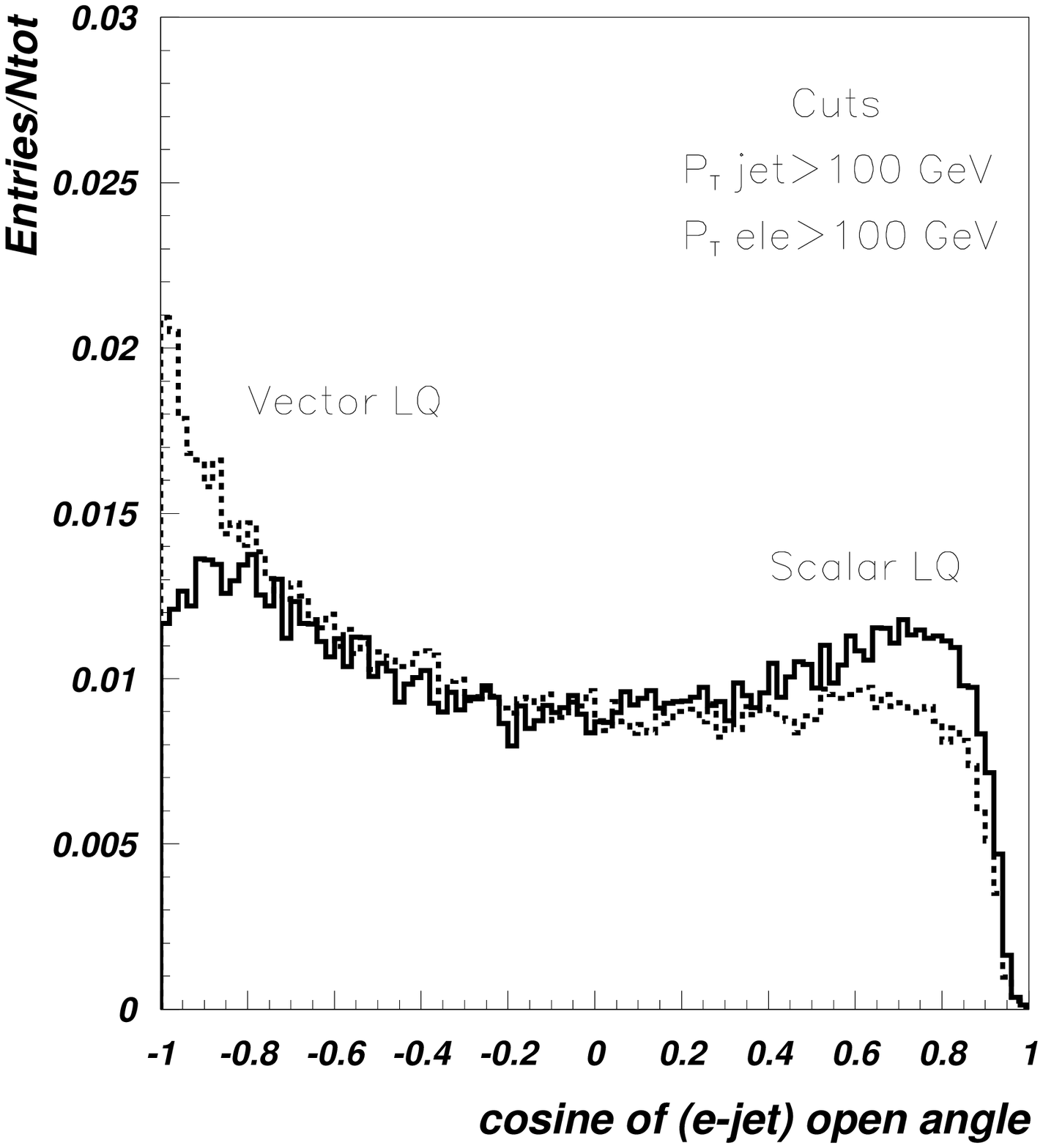}
\caption{Distributions of the cosine of the open angle between electron and jet in
events with a scalar~(solid line) and  vector~(dashed line)  $\LQ$ single production.
\label{fig2}}}  
Besides the difference in cross section
for vector and scalar $\LQ$ production 
one could expect a difference in angular correlations
of  vector $\LQ$ decay products 
compared to scalar $\LQ$ if the vector $\LQ$
is being produced with some polarization.
We have found that,  indeed, vector $\LQ$
is  produced with non-trivial polarization at the LHC
which is reflected, for instance,
in a  difference  in the angular distribution
between  the leading \pT electron and jet
shown as an example in Fig.~\ref{fig2} for $\LQ$ single production case.
The study of such angular correlation effects 
is, however,  beyond the scope of the present article.
The distributions of kinematical characteristics we have chosen in this study
are similar for  scalar and vector $\LQ$ production.

The ATLFAST\cite{19} code has been
used to take into account the experimental conditions prevailing at LHC for the 
ATLAS detector. The detector concept and its physics potential have been presented
in the Technical Proposal\cite{20} and the Technical Design Report\cite{21}.  The
ATLFAST code for fast detector simulations accounts for most of the detector
features: jet reconstruction in the calorimeters, momentum/energy smearing for
leptons and photons, magnetic field effects and missing transverse energy. It
provides a list of reconstructed  jets, isolated leptons and photons. In most
cases, the detector--dependent parameters were tuned to values expected for the
performance of the ATLAS detector from full simulation.

The electromagnetic calorimeters were used to reconstruct the energy of leptons
in cells of dimensions $ \Delta \eta \X \Delta \phi =0.025\X 0.025 $
within the pseudorapidity  ($ \eta $) range $ -2.5<\eta <2.5 $; $\phi$ is the azimuthal 
angle. 
The electromagnetic energy resolution is given by $ 0.1/\sqrt{E}(GeV)\bigoplus 0.007 $ 
over this pseudorapidity region.
The electromagnetic showers are identified as leptons when they lie within a cone
of radius $ \Delta R=\sqrt{(\Delta \eta)^{2} \X (\Delta \phi)^{2}} = 0.2 $ 
and possess a transverse energy $ E_{T}>5 $ GeV. 
Lepton isolation criteria were applied, requiring a distance $\Delta R>0.4$
from other clusters and maximum transverse energy deposition, $ E_{T}<10 $ GeV, 
in cells in a cone of radius $ \Delta R=0.2 $ around the direction of electron emission.

Jet energies were reconstructed by clustering hadronic calorimeters cells
of  dimensions $ \Delta \eta \times \Delta \phi =0.1\times 0.1 $
within the pseudorapidity range $ -2.5<\eta <2.5 $.  
The hadronic energy resolution of the ATLAS detector is 
parametrized as $0.5/\sqrt{E(GeV)}{\oplus }0.03 $  over this $ \eta  $ region. 
Hadronic showers are regarded as jets if the deposited transverse energy $ E_{T} $ 
is greater than 15 GeV within a cone of radius $\Delta R=0.4$.

It must be mentioned that standard parametrization in ATLFAST has been used
for the electron resolution but detailed studies are needed, using test beam 
data and GEANT full simulation to validate the extrapolation of the resolution function 
to electron energies in the TeV range.

In this paper, two types of signal event have been studied.

\begin{enumerate}
\item {\bf Type 1}($2l+jets$) signature, for which 
$\LQ$ produced singly in association with an electron,  decays to an electron and a quark,
while each $\LQ$ produced in pair is 
required to decay to an electron and a quark.    

\item {\bf Type 2}($l+jets+E_T^{miss}$) signature, for which 
$\LQ$ produced singly in association with a neutrino,  decays to 
an electron and a quark.~\footnote{The same signature will be for single $\LQ$
production in association with the electron if  $\LQ$ decays into neutrino and quark.}
For $\LQ$ pair production, 
one  $\LQ$ is required to decay into electron and quark, while the other one --  to a neutrino 
and a quark.  
\end{enumerate}

Backgrounds of Type~1 signal signature are $Z+jets$ events, where $Z$ decays into two
electrons, and $\ttb$ events where both $W$ from the top quark decay into an electron
(positron) and a neutrino. For the signal events of Type~2,  backgrounds are
$W+jets$ events, where $W$ decays into an electron and a neutrino, and $\ttb$ events
where one $W$ decays into an electron and neutrino, and the second $W$ decays into jets.

\TABLE{
\caption{\label{table2}Cross section $\sigma$(pb) from PYTHIA for the backgrounds after pre-selection cuts.}
\begin{tabular}{ccccc} \hline
\hline
Process      &$\ttb$(Type 1)& $ \ttb$ (Type 2) & $ Z+jet $ & $  W+jets $\\ \hline
$\sigma$~(pb)&   6.6         &        11       &     665   &     17.    \\ \hline
\end{tabular}
}

In order to enrich the event statistics in the region of high invariant
masses, simulated background events were pre-selected in PYTHIA for hard $2\rightarrow2$ 
process with transverse momentum, $\hat{p}_T> 200$ GeV (100 GeV for $Z+jet$ events), 
defined in the rest frame of the hard interaction, followed by the standard initial and 
final state radiation technique. Top quark pair production
and $W+jets$ background events for Type~2 signal events were additionally
pre-selected with at least one electron and $\eslt >100$ GeV. Corresponding
cross sections for the backgrounds, used for this article, are shown in 
Table~~\ref{table2} for the events, passing pre-selection cuts.
Since the   signal events 
would consist of events originated from the single or pair production of leptoquarks, 
we tried to define a unique set of cuts effective for the background suppression 
against combined $\LQ$  pair+single signal.

\subsection{Type 1 $\LQ$ signal events}

Here, we present the analysis of $\LQ$ events of Type 1. The signal signatures consist of at least one jet and two electrons.

Due to their large mass, leptoquarks would produce events with large transverse energy.
The scalar sum of transverse momenta of all charged particles in the event, $H_{T}$, is presented for single, 
Fig.~\ref{fig3}(left), and pair, Fig.~\ref{fig3}(right), scalar $\LQ$ production.
As can be seen, an appropriate choice of a $H_{T}$ value should suppress 
the bulk of background events.

\FIGURE{
\includegraphics[width=7.8cm,height=7.8cm]{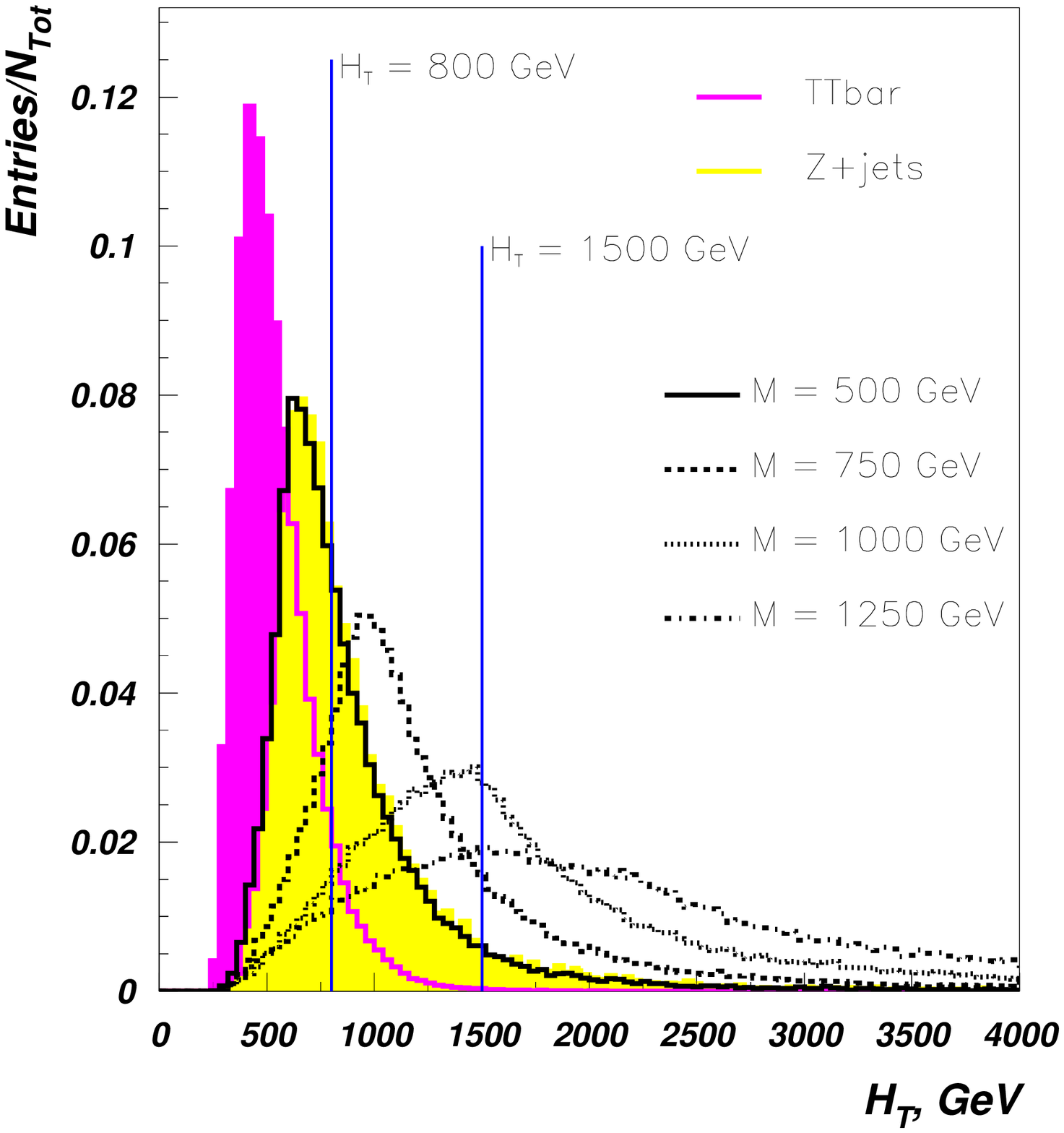}%
\includegraphics[width=7.8cm,height=7.8cm]{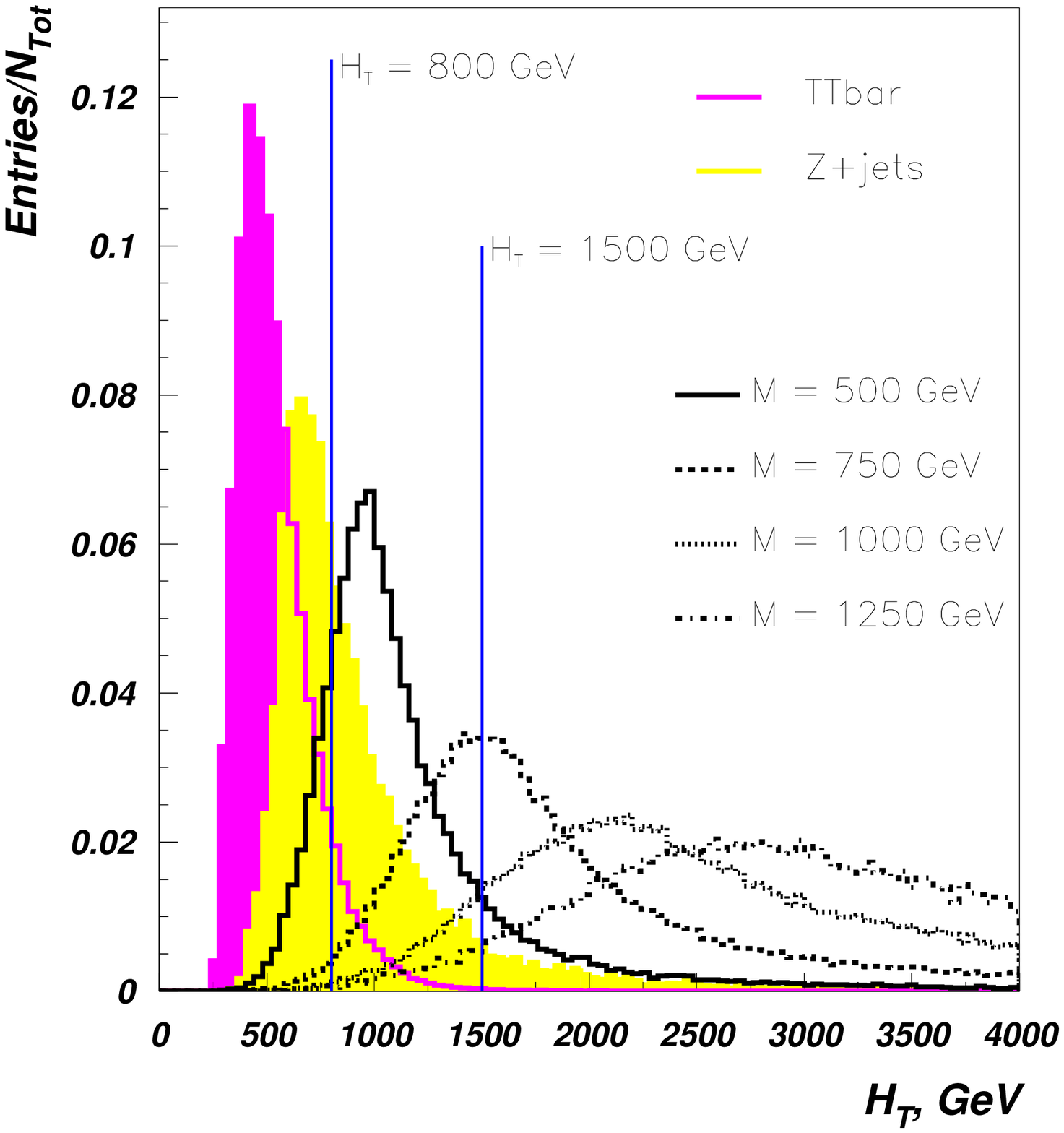}
\caption{ Distribution of the $H_{T}$ variable (see text) for
events of single (left) and pair (right) production of scalar 
leptoquarks and corresponding backgrounds. All distributions are normalized to unity.
\label{fig3}}}

The following cuts were used to separate signal from background:

\begin{itemize}

\item  The transverse momentum of two electrons were required to be at least
       90 GeV (100 GeV in the case of $M_\LQ>750$ GeV). 
\item  At least one jet was required with minimum transverse momenta of 70 GeV 
       (90 GeV in the case of  $M_\LQ=750$ GeV and $>100$ GeV/c for higher $\LQ$ masses).
\item  The invariant mass of two electrons was required to be larger than 
150 GeV in order to veto the dominant $Z+jets$ background.
\item  Events with at least one b-jet were vetoed to suppress $\ttb$ background.
\item  The scalar transverse momentum sum of all selected particles in the 
event, $H_{T}$, was required to be at least 800 GeV.

\end{itemize}
 \FIGURE{
\includegraphics[width=7.8cm,height=7.8cm]{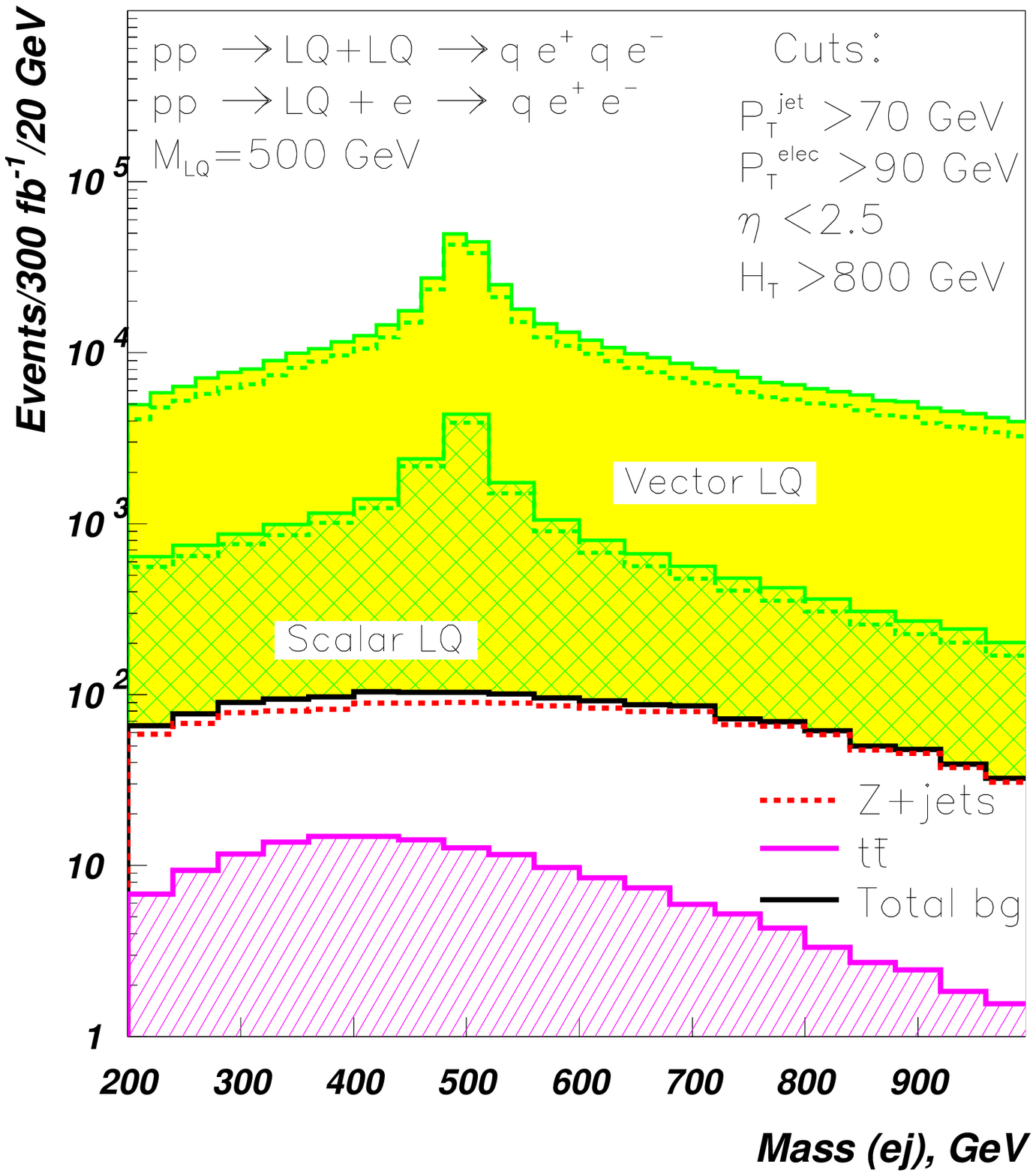}%
\includegraphics[width=7.8cm,height=7.8cm]{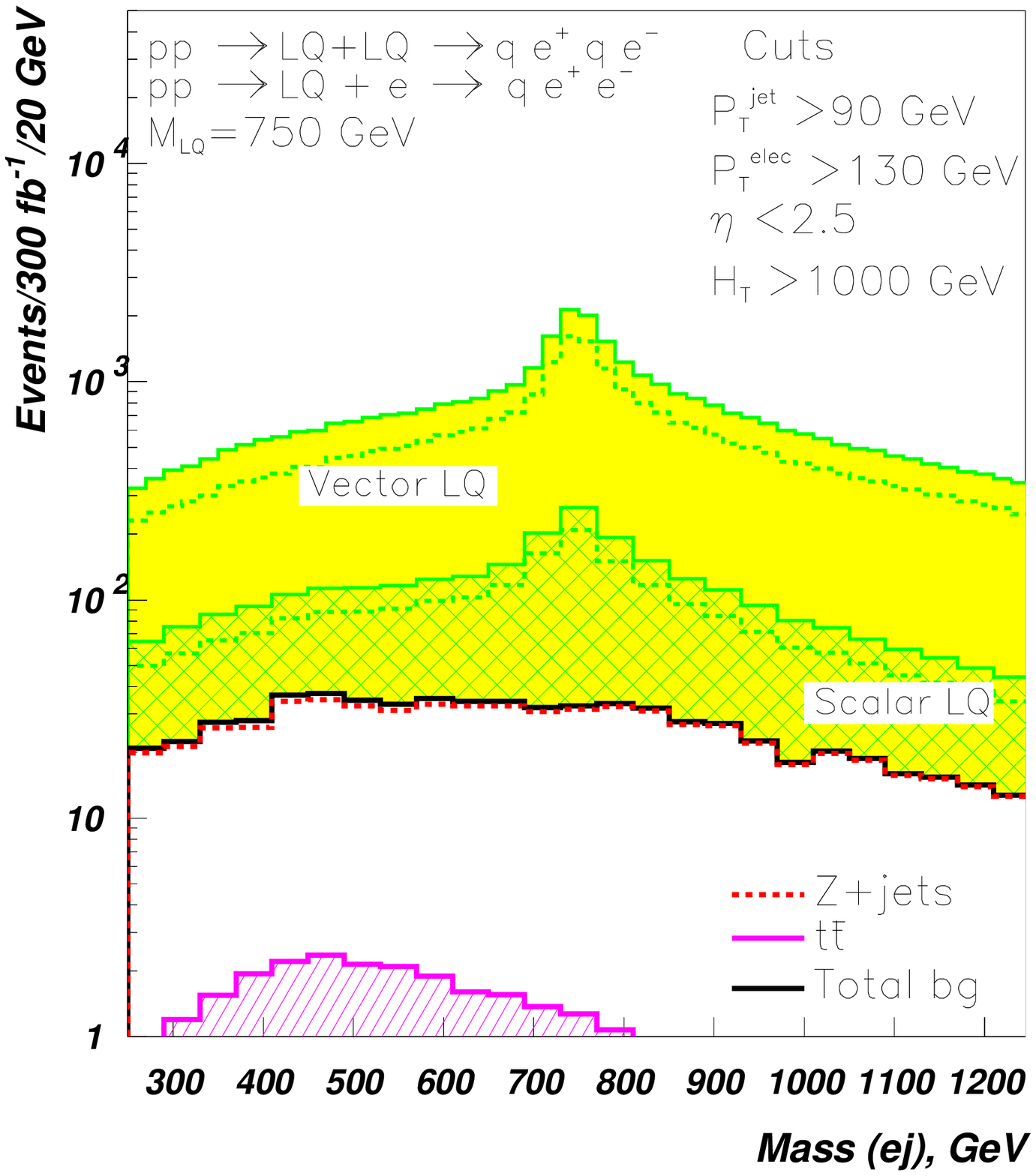}
\includegraphics[width=7.8cm,height=7.8cm]{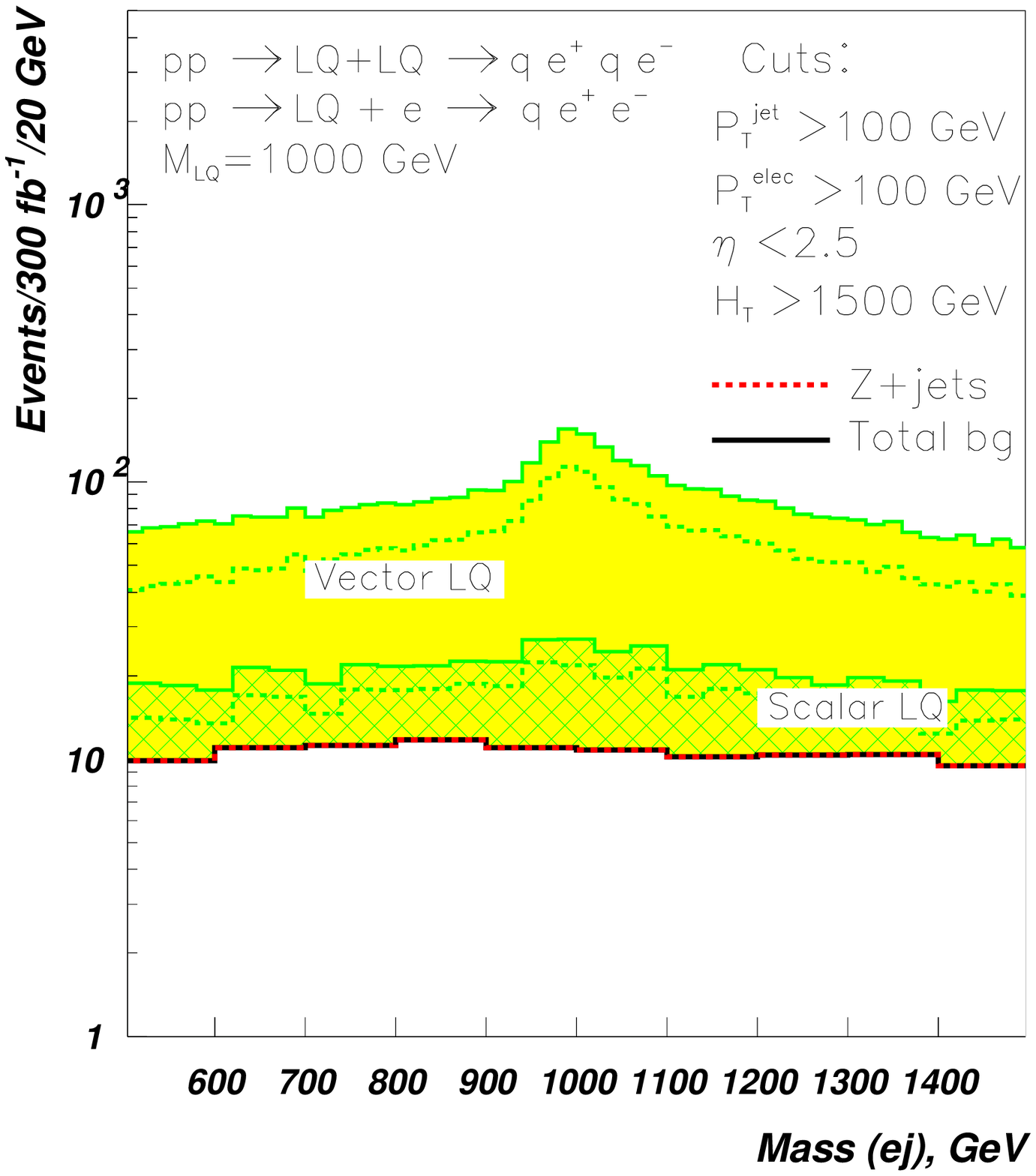}%
\includegraphics[width=7.8cm,height=7.8cm]{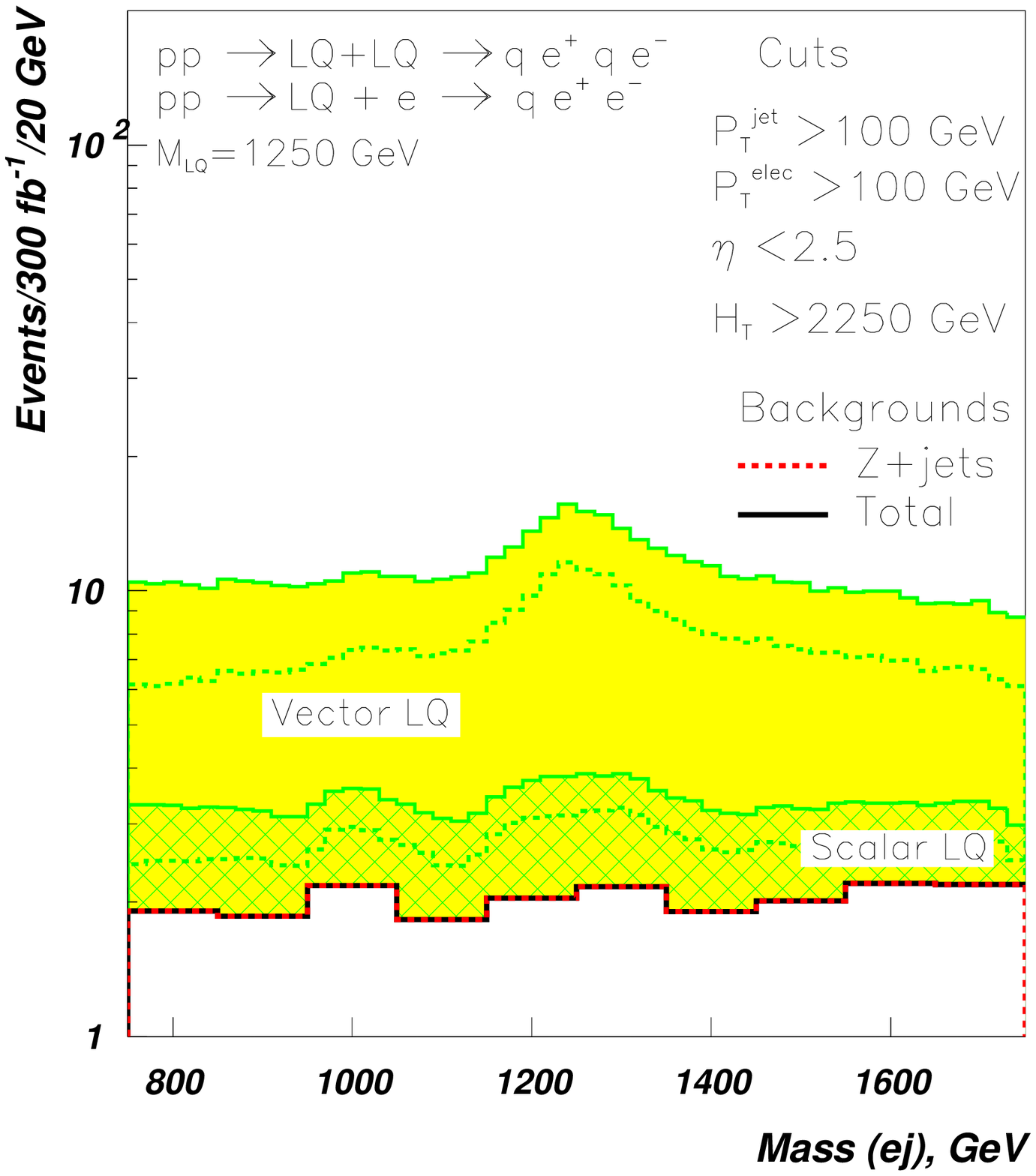}
\vspace*{-0.5cm}
\caption{Invariant mass distribution of electron and jet for events of single 
and pair production of scalar and vector leptoquarks mass $500,750,1000$ and $1250$~GeV 
for  Type 1 signal events. 
\label{fig5}}}

     The resulting invariant mass distributions for the electron-jet system are 
presented in Figs.~\ref{fig5} for combined single  and  pair 
production of leptoquarks of various masses. Signal distributions are presented 
for scalar (hatched area) and vector leptoquarks. 
The dashed line shows the pair production contribution to the total signal 
spectrum. The contribution of the single production to the total spectrum gradually increases with the $\LQ$ mass.  
The cross section  for vector leptoquarks corresponds to MC case ($\kappa_G=1, \lambda_{G}=0$).

All possible mass combinations between two leading \pT jets and electrons are allowed, which leads to some broadening of the signal distributions for the single production case. 

As can be seen from these figures, the $Z+jets$ background is essentially dominant.
The pair production of top quarks gives a rather small contribution and can be 
seen only for the lowest leptoquark mass case studied.

The signal statistical significances are given in Table ~\ref{table3}.
The data are presented for combined~(single+$\LQ$ pair production)  signal efficiencies,
with the number of events shown for the total background and for
$\LQ$ single and pair  production  for different masses of scalar and vector leptoquarks.
The data are presented for an integrated luminosity of $L=300 fb^{-1}$.

\TABLE{
\begin{tabular}{c|r| rrr||rrr} 
\hline \hline
$\LQ$ Mass &  Background &\multicolumn{3}{|c|}{Scalar $\LQ$} & \multicolumn{3}{|c}{Vector $\LQ$}\\
\cline{3-8}
(GeV) &     & $\LQ$-Single  & $\LQ$-Pair 	& $S_{tot}/\sqrt{B}$  & $\LQ$-Single 	& $\LQ$-Pair   	& $S_{tot}/\sqrt{B}$\\
\hline
500   & 614 & 1835       & 14529   &    660     & 25907  & 155374    & 7316   \\
750   & 264 & 335	 & 1033	   &     84     &  2902	 &  8571     &  706   \\
1000  & 132 &  54	 & 112	   &     14     &   392	 &   894     &  112   \\
1250  &  40 &  12	 & 17	   &      5     &    70	 &   140     &   33   \\
1500  &  28 &   4	 & 3	   &      1     &    16	 &    20     &    7   \\
\hline \hline
\end{tabular}
\caption{
Number of signal events for $\LQ$ single and pair production versus total
background and respective significance $S_{tot}/\sqrt{B}$ for combined signal ($\LQ$
 pair + single production). Results are for an integrated luminosity of $L=300 fb^{-1}$ and
Type~1 signal  signature. \label{table3}}
}

\subsection{Type 2 $\LQ$ signal events}

In this section we present the analysis of signal events with an electron, jets and a neutrino.
The signal signature is at least one jet, electron and missing transverse momenta.
The scalar sum of transverse momenta of all charged particles in the event, 
$H_{T}$, is presented for single, 
Fig.~\ref{fig7}~(left), and pair production, Fig.~\ref{fig7}~(right) of scalar $\LQ$.
As can be seen, an appropriate choice of a $H_{T}$ value again should suppress 
the bulk of background events.  
The presence of a neutrino in the $\LQ$ signal events of this type, provides
a possibility to suppress relevant backgrounds using a missing transverse momentum 
variable.
In Fig.~\ref{fig8}~(right), we plot the orbital ($\phi$)-angle difference, $\Delta\phi$, 
between electron and the missing transverse momentum vector. Distributions are shown for 
scalar and vector leptoquark decays and corresponding backgrounds.
The background spectra reveal the emission of an electron and a neutrino along the 
same direction, which is reasonable, since in background events those particles are 
produced in $W$ decays. The signal distributions bear mostly the back-to-back emission 
feature.
\FIGURE{
\includegraphics[width=7.8cm,height=7.8cm]{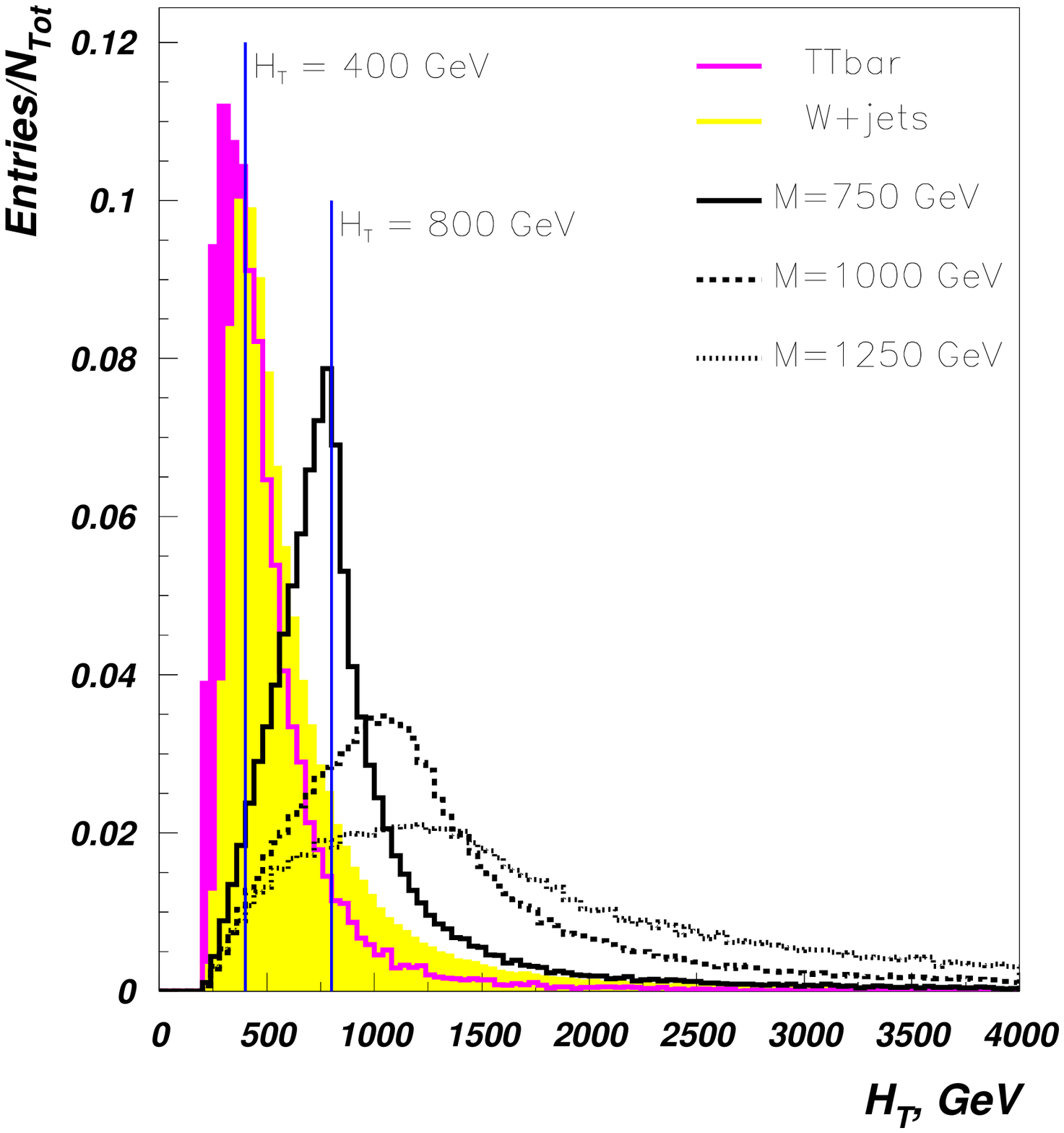}%
\includegraphics[width=7.8cm,height=7.8cm]{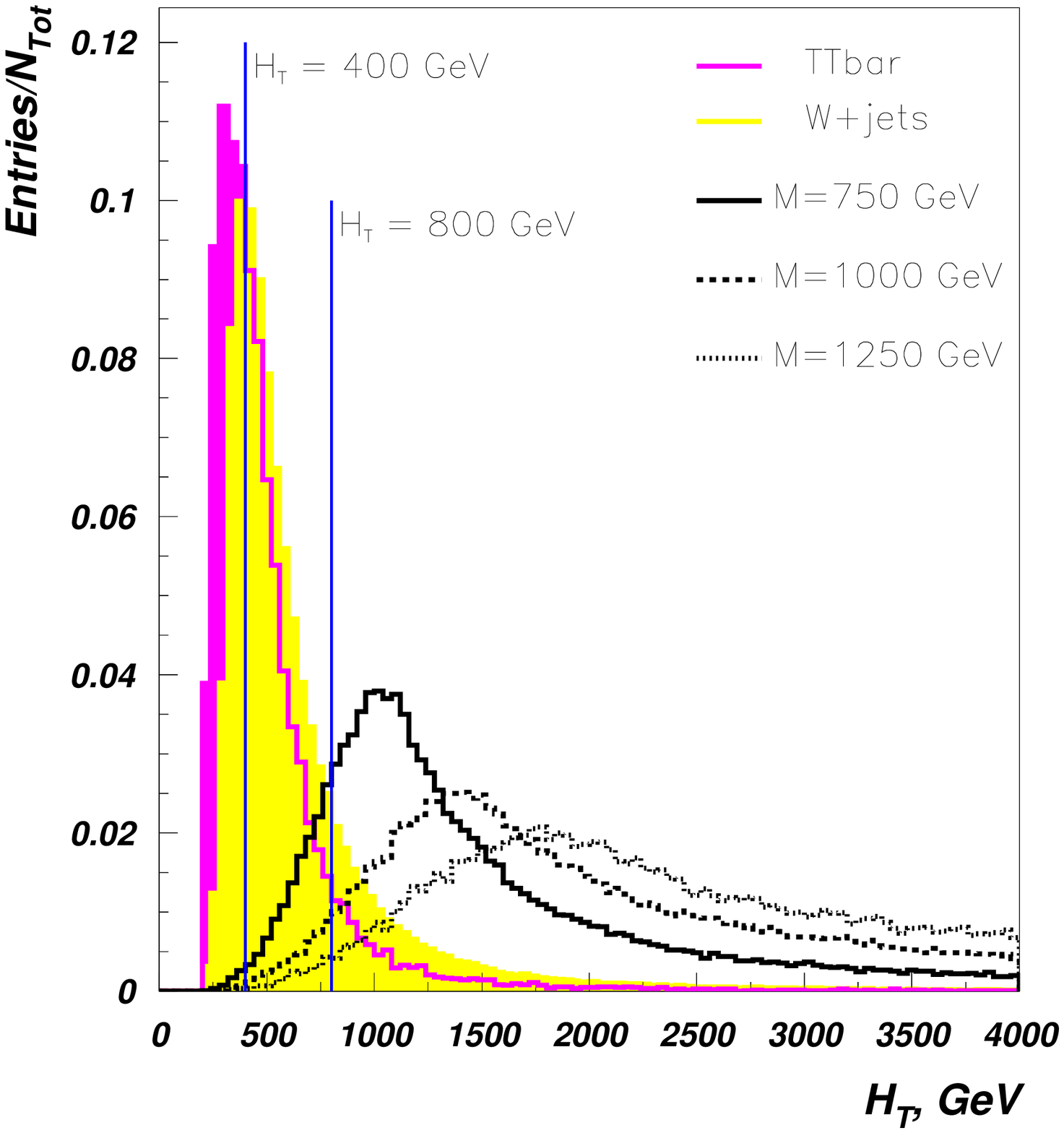}
\vspace*{-0.9cm}
\caption{ Distribution of $H_{T}$ variable (see text) for
events of single (left) and pair (right) production of
 leptoquarks and corresponding backgrounds.
  Decays of $\LQ$ include electrons and neutrino.  All distributions are normalized 
to unity. 
\label{fig7}}}
The following set of cuts was worked out  to effectively  separate the signal from background:
\begin{itemize}

\item  The transverse momentum of the electron was required to be at least
       100 GeV.
\item  At least one jet was required with a minimum transverse momentum of at least 
       100 GeV.
\item  The transverse mass of an electron and the missing transverse momentum vector 
       ($P_T^{miss}$, see Fig.~\ref{fig9}) was required to be larger than 200 GeV in 
       order to veto the dominant $W+jets$ background.
\item  Events with at least one b-jet were vetoed to suppress $\ttb$ background.
\item  The scalar sum of transverse momenta of all selected particles in the
       event, $H_{T}$,  was required to be at least 400,(600, 800, 1000, 1200) GeV
       for $\LQ$ masses of 500, (750, 1000, 1250, 1500) GeV, respectively.
\item  The orbital angle difference between an electron and the transverse missing
       momentum vector was required to be greater than 0.8 radians.
\end{itemize}
\FIGURE{
\includegraphics[width=7.8cm,height=7.8cm]{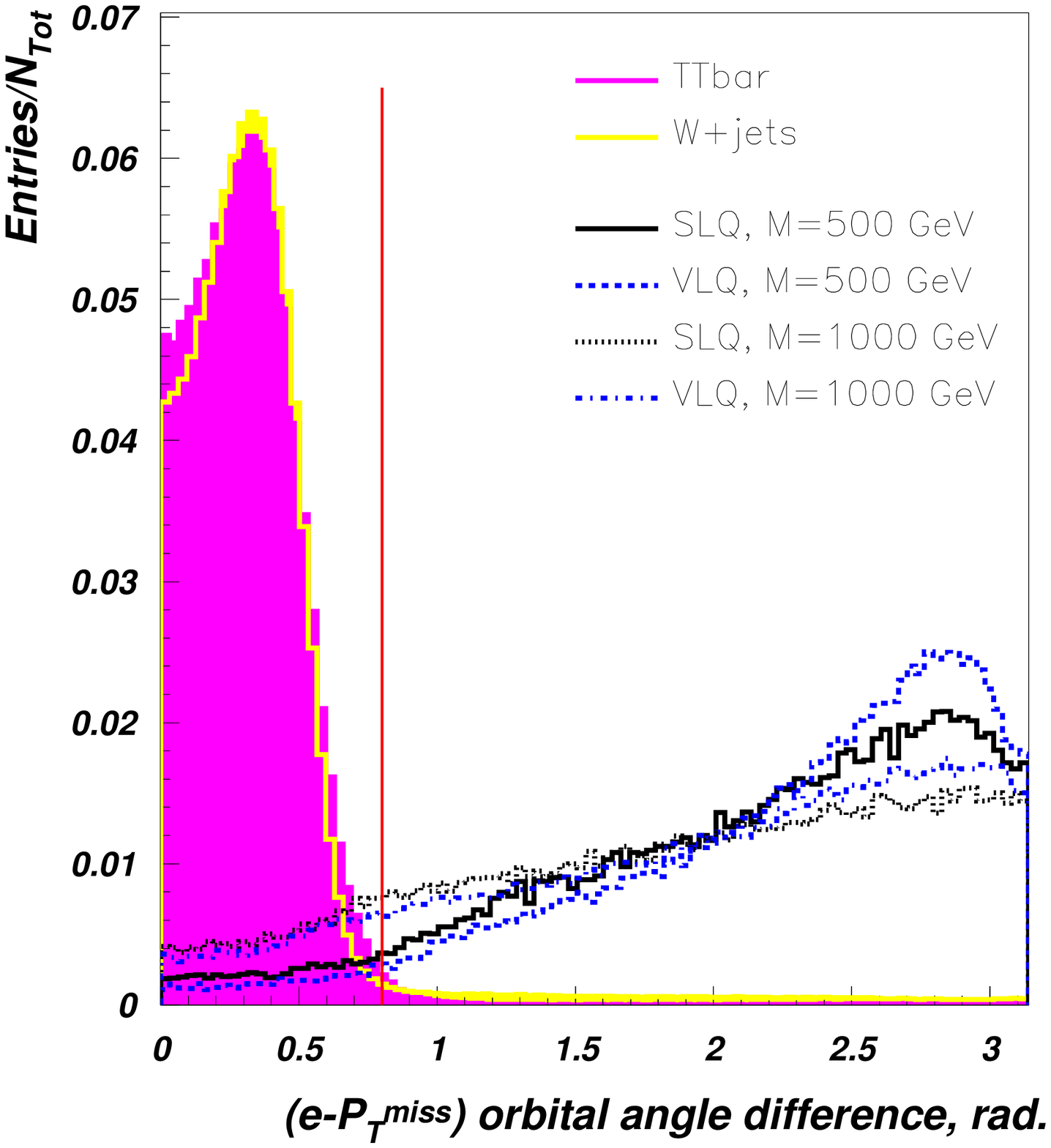}%
\includegraphics[width=7.8cm,height=7.8cm]{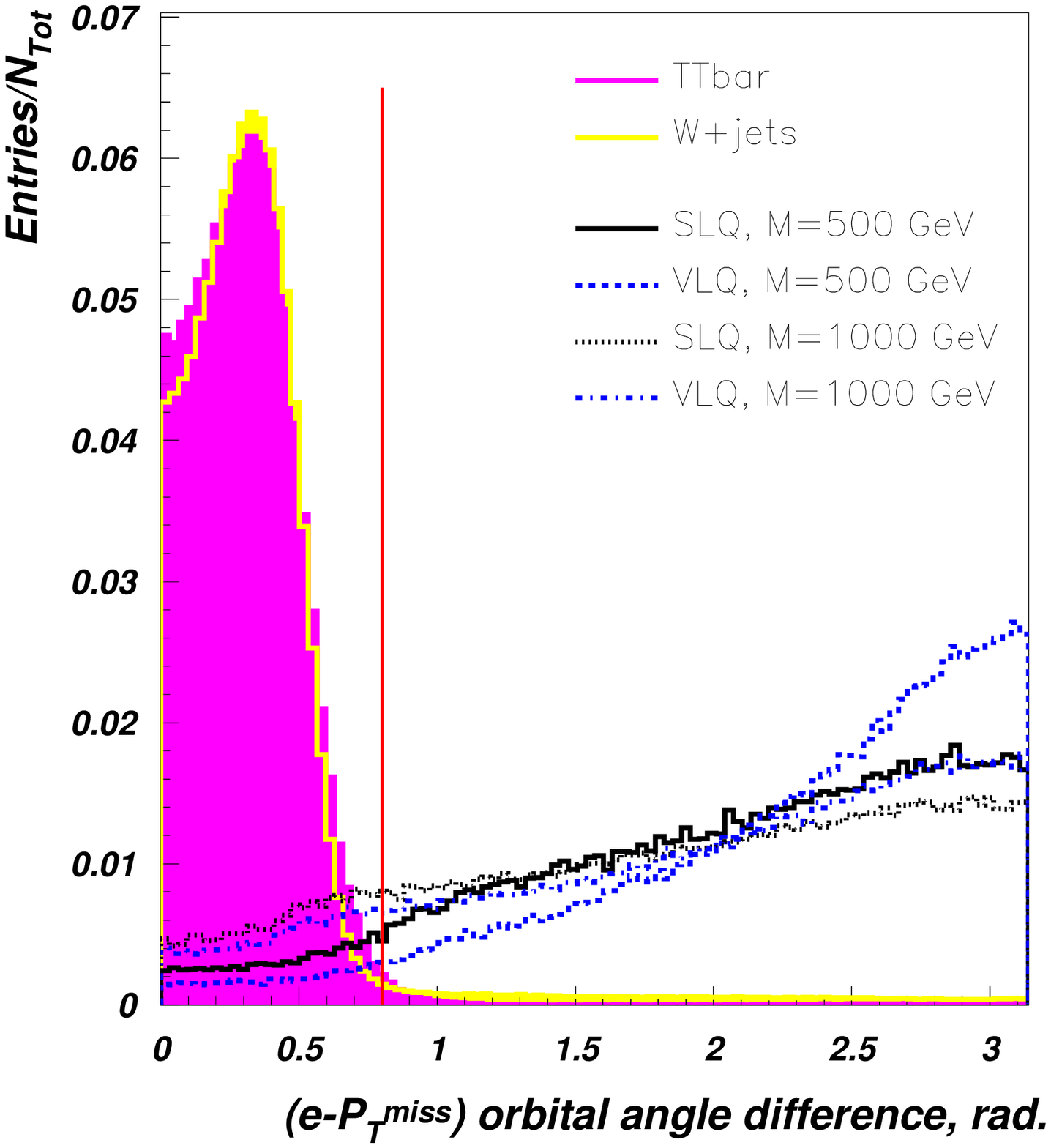}
\vspace*{-0.9cm}
\caption{The distribution of the orbital $\phi$-angle difference between an electron and 
the missing transverse momentum vector, $\Delta\phi$, in radians, for signal (scalar and vector $\LQ$) and background $W+jets$ and $\ttb$ events.
single~(left) and pair~(right) $\LQ$ production are presented.    
\label{fig8}
}}
While the angular spectra of scalar and vector $\LQ$ of $500$ GeV mass are quite close to 
the background spectra in the forward hemisphere, the signal distributions for larger
masses allow the separation of the signal from the background.  
     The resulting invariant mass distributions for the electron-jet system of the signal
 events of Type 2 are presented in Fig.~\ref{fig10} for the $\LQ$ mass of 500 GeV 
(left side) and  750 GeV~(right side). 
Signal distributions are presented for scalar leptoquarks (hatched
 area) and vector leptoquarks for the minimal coupling set. 
The dashed line shows the pair production contribution to the total signal 
spectrum. Similarly to the Type 1 signal events, the contribution of the single 
production to the total spectrum gradually increases with the $\LQ$ mass.
The signal statistical significances are reported in Table ~\ref{table4}
combined for $\LQ$ single and pair production
for different $\LQ$ masses. Table ~\ref{table4} also presents 
number of events, separately for  for $\LQ$ single and pair production
as well as for background events. 
The data correspond to an integrated luminosity of $L=300 fb^{-1}$.
\FIGURE{
\includegraphics[width=7.8cm,height=7.8cm]{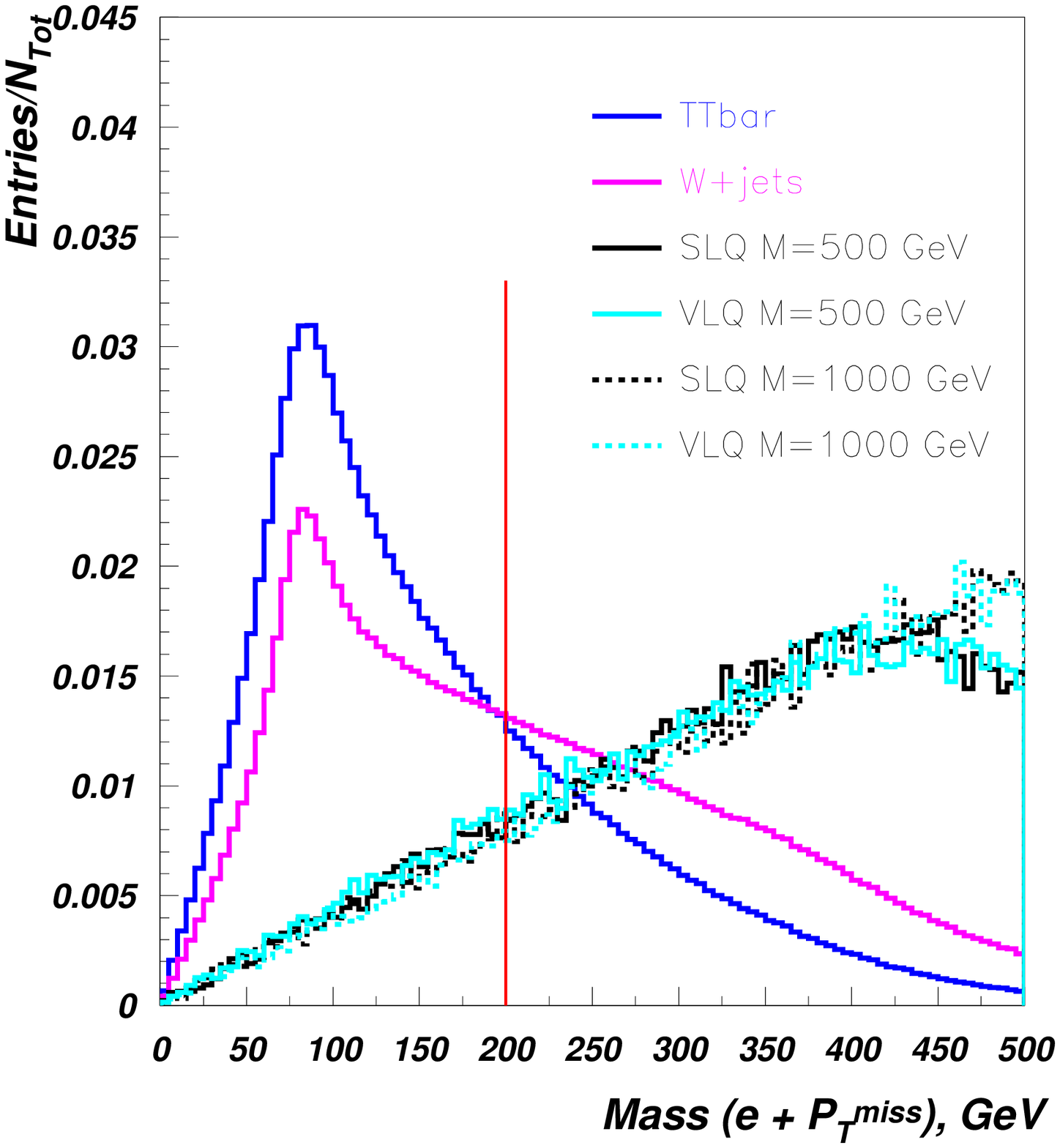}%
\includegraphics[width=7.8cm,height=7.8cm]{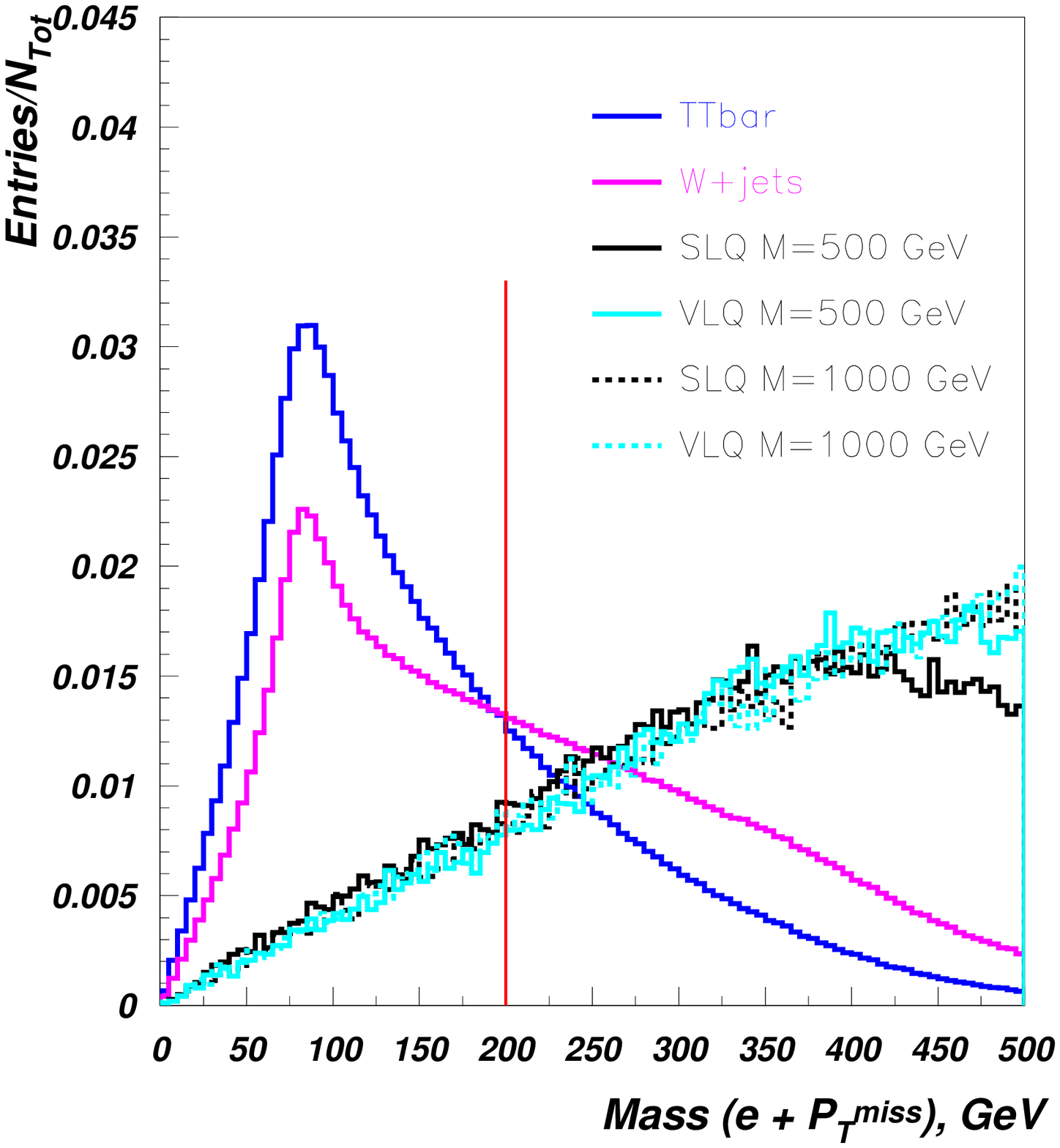}
\caption{The distribution of the transverse mass of an electron and the $P_T^{miss}$ 
vector for signal (scalar and vector $\LQ$) and  $W+jets$ and $\ttb$ backgrounds.
single~(left) and pair~(right) $\LQ$ production are presented.
\label{fig9}
}
}

\FIGURE{
\includegraphics[width=7.8cm,height=7.8cm]{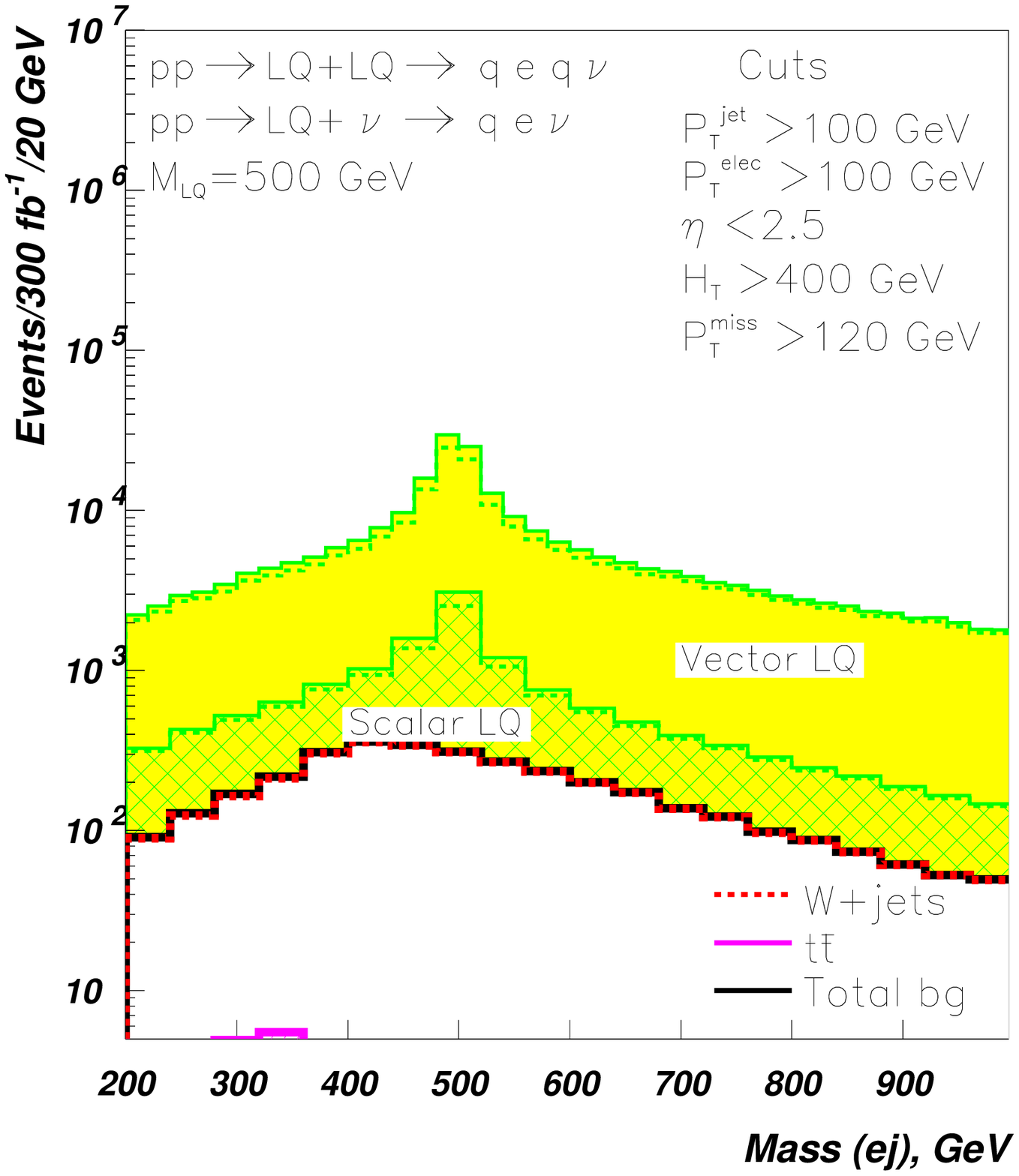}%
\includegraphics[width=7.8cm,height=7.8cm]{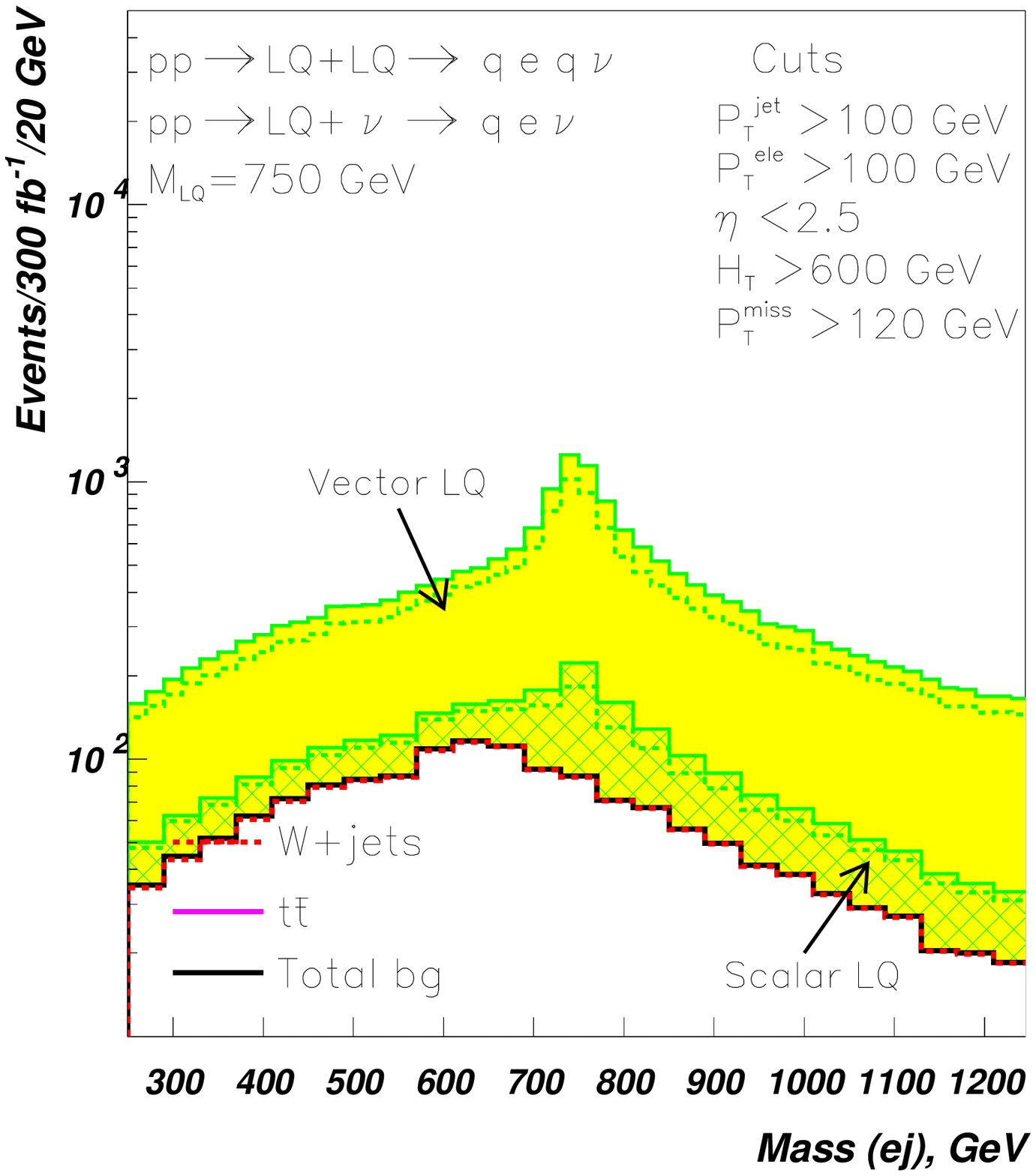}
\includegraphics[width=7.8cm,height=7.8cm]{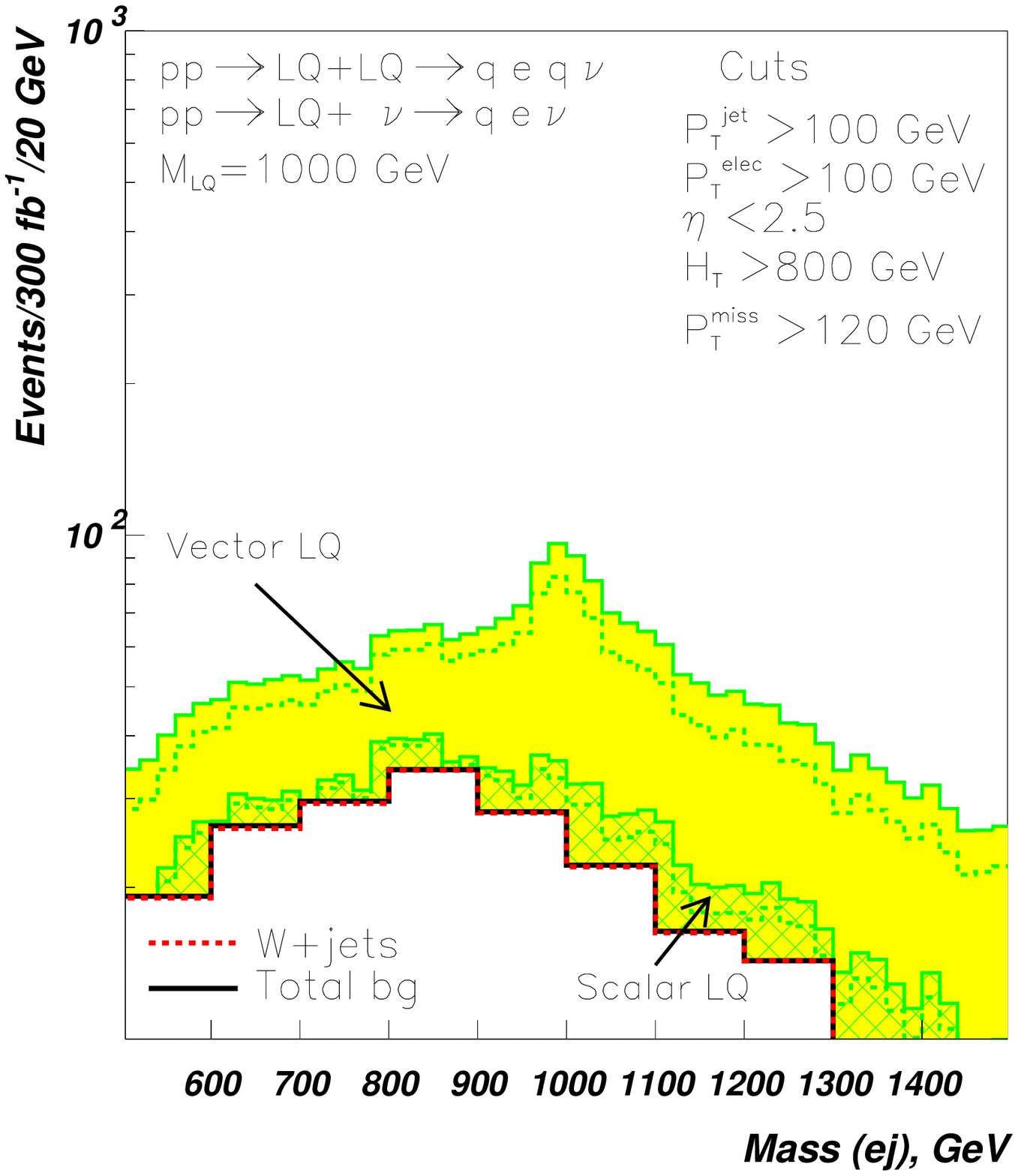}%
\caption{Invariant mass distribution of electron and jet for events of single  
and pair production of scalar and vector leptoquarks for $m=500$ GeV (upper left) and $m=750$~GeV (upper right) and $m=1000$ GeV (below) for Type 2 signature. 
\label{fig10}}}

\TABLE{
\begin{tabular}{c|r| rrr||rrr} 
\hline \hline
Mass &  Background &\multicolumn{3}{|c|}{Scalar $\LQ$} & \multicolumn{3}{|c}{Vector $\LQ$}\\
\cline{3-8}
(GeV) &     & $\LQ$-Single  & $\LQ$-Pair  & $S_{tot}/\sqrt{B}$	& $\LQ$-Single 	& $\LQ$-Pair   	& $S_{tot}/\sqrt{B}$\\
\hline
500   & 1850 &  2093     &   8383   &      244 &  17889    &  88438  	&  2425      \\
750   &  674 &   251	 &    545   &       31 &   1229    &   4953	&   238      \\
1000  &  303 &    33	 &     52   &        5 &    135    &    460	&    34      \\
1250  &  149 &     5	 &      7   &        1 &     17    &     64	&     7      \\
\hline \hline
\end{tabular}
\caption{Number of signal events for single  and $\LQ$ pair production versus total background
and respective significance $S_{tot}/\sqrt{B}$
for combined signal (pair + $\LQ$ single production).
Results are for an integrated luminosity of $L=300 fb^{-1}$ and  Type 2  signature.
\label{table4}
}
}

\subsection{$\LQ$ mass reach of LHC}

The $\LQ$ mass reach of  LHC is shown in Fig.~\ref{fig11} for
combined single  and $\LQ$ pair production processes and two types of signal 
signatures.
The results for scalar and vector $\LQ$ are presented separately. 
The data are presented for an integrated luminosity of $L=300 fb^{-1}$.
One can see that scalar leptoquarks can be accessible at LHC up to masses $\lesssim 1.2$~TeV 
while vector $\LQ$ for MC case can be discovered for masses  $\lesssim 1.5$~TeV.
LHC can exclude  $\LQ$ at $95\%$~CL  with masses about 200~GeV
above  $5\sigma$ discovery $\LQ$ limit, i.e.  with masses  $\lesssim 1.4$~TeV  and $\lesssim 1.7$~TeV 
for scalar and vector~(MC) $\LQ$, respectively.

Let us remind, that we have chosen $\LQ-l-q$ coupling $\lambda_{eff}=e$ and
contribution from $\LQ$ single production rescales quadratically with this coupling.
For other values of  $\lambda_{eff}$, the  new LHC reach can be easily  found
by using LHC reach Tables~5 and 6.
One can see that Type~1 signature ($2\ell+jets$) looks more promising
compared to Type~2 events ($2\ell+jets+\eslt$).

\FIGURE{
\includegraphics[height=10.5cm]{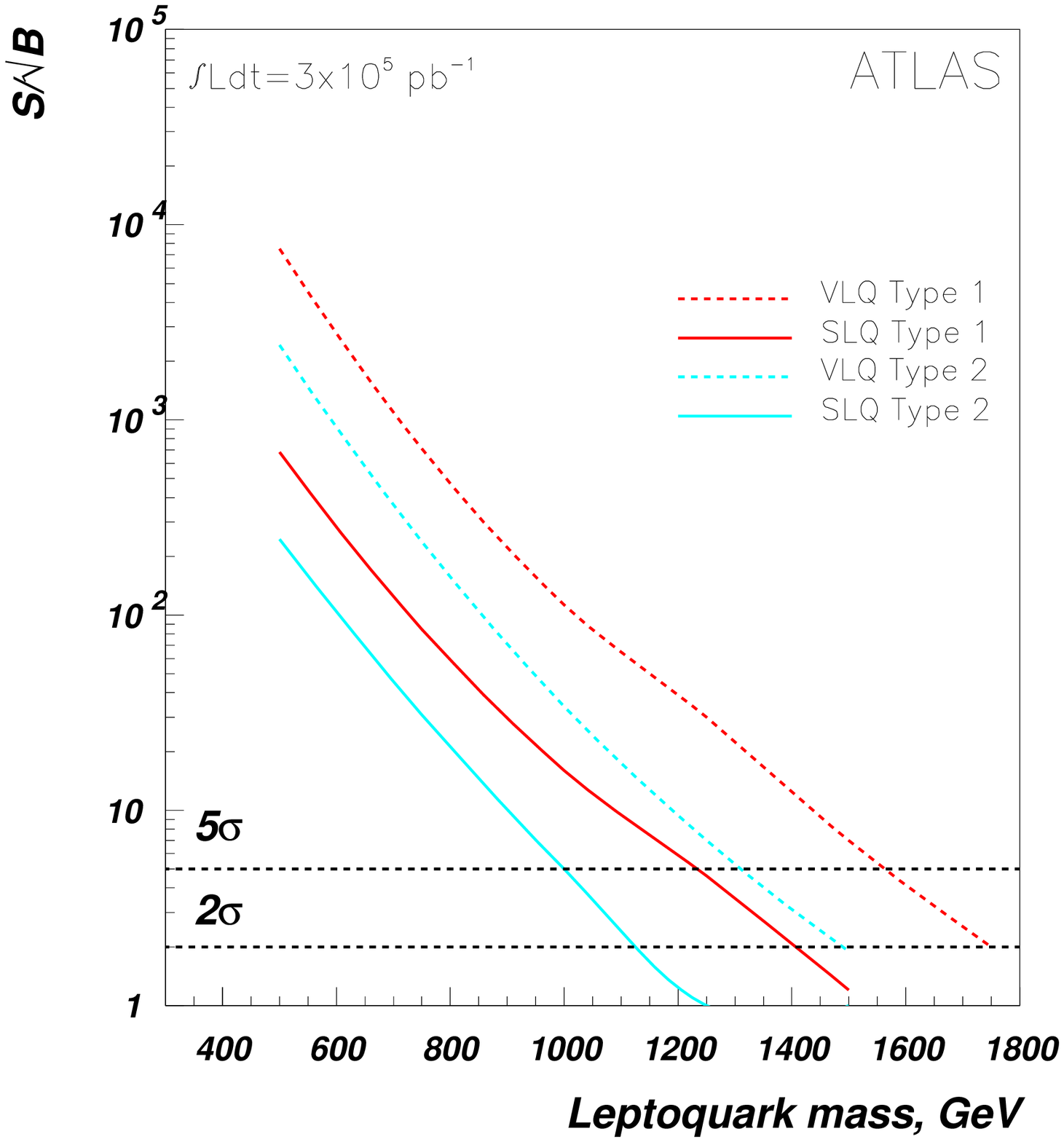}
\caption{Leptoquark mass reach of the LHC at $5\sigma$~level for Type~1 and Type~2
signal signatures for scalar (solid line) and vector (dashed line) $\LQ$
for an integrated luminosity of $L=300 fb^{-1}$.
For vector  $\LQ$, the MC choice has been selected. 
The contributions of single  and pair $\LQ$ signal are combined.
\label{fig11}
}
}
One should  notice that, for the chosen $\lambda_{eff}$, the contribution from 
 $\LQ$ single production is  30-50\%
to the total number of signal events for the $\LQ$ mass at the discovery limit.
Therefore, $\LQ$ single and pair production 
should be studied together at the LHC~\footnote{
Eventually, the case of single
production of $\LQ$ of the second and third generations
is qualitatively different. In this case only pair $\LQ$
would give the major contribution to the signal rates,
unless $\lambda_{eff}$ is too large (which might be not 
allowed by other experimental constraints) 
to enhance $\LQ$ single production, suppressed due to  initial sea-quarks PDFs.}.
The complementarity of single  and pair $\LQ$ channels
is clearly illustrated in our final Fig.~14, where we present
scalar and vector leptoquark mass reach of the LHC in the  
($M_\LQ - \lambda_{eff}$) plane. For example,
for $\lambda_{eff}=1$, $\LQ$ single production
allows the extension of  LHC reach for  $M_\LQ$ by about 400 GeV!

\FIGURE{
\includegraphics[height=16.5cm]{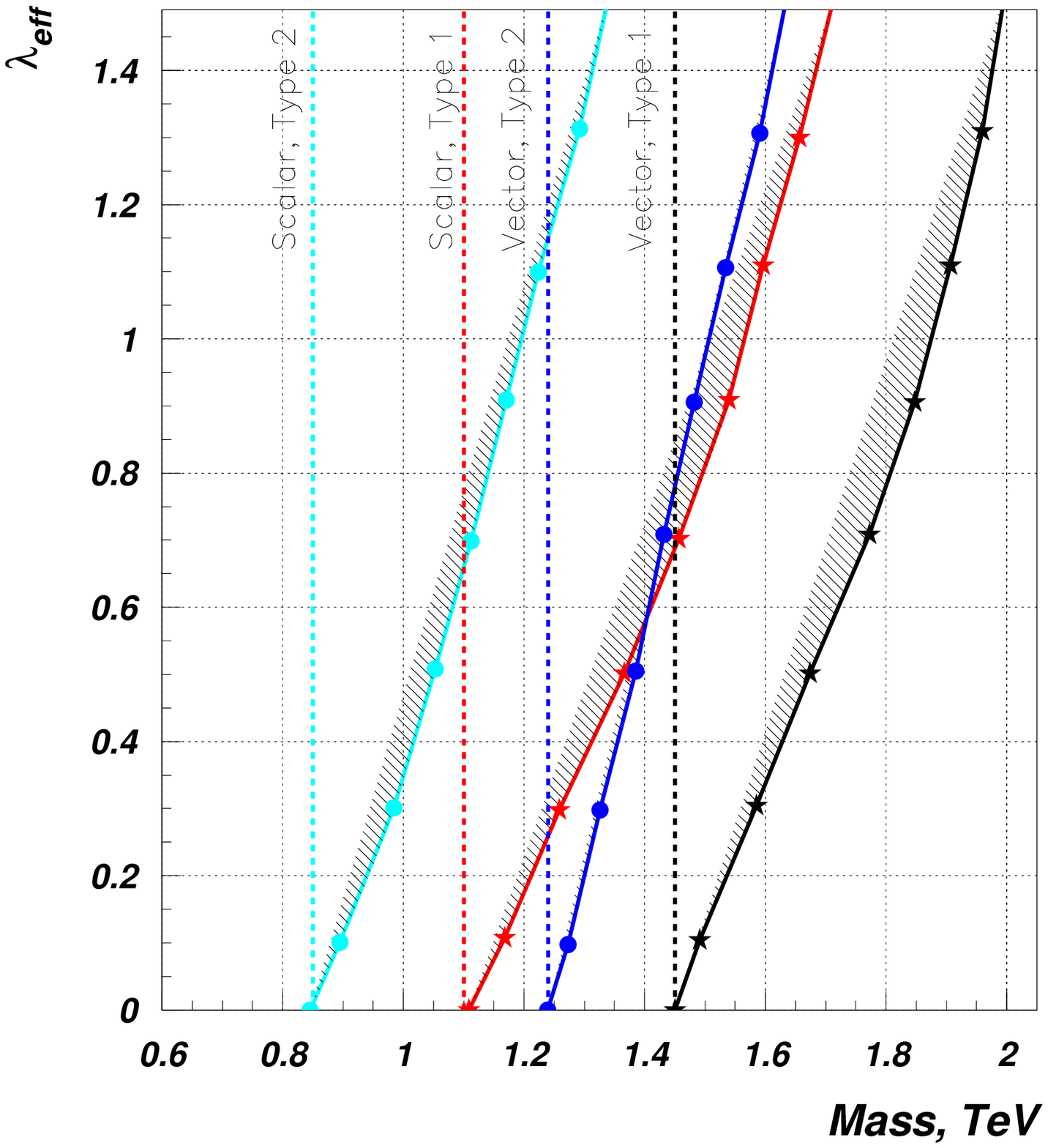}
\caption{Scalar and vector leptoquark mass reach of the LHC for 
($M_\LQ -- \lambda$) plane
for $\LQ$ pair production~(dashed line) and for combined
$\LQ$ single and pair production~(solid line).
Type 1 signal  reach contour is denoted by stars while
Type 2 signal  reach contour is marked by solid dots.
\label{fig12}}}

\section{Conclusions}

In this paper, we present a new detailed  study of  $\LQ$ production and decay 
at  LHC at the level of detector simulation.

We treated  $\LQ$ single and pair production together 
and have worked out a set of kinematical cuts to maximize
significance for combined single and pair production events.

It was  shown, that 
combination of signatures from  $\LQ$ single and pair production
not only significantly increases the LHC reach, but also 
allows us to give the correct  signal interpretation.
Our results are summarized in Tables~5 and 6 and
Figs.~14 and 15.  In particular, the LHC can
discover $\LQ$ with a mass up to  $1.2$~TeV  and $1.5$~TeV 
for the case of scalar and vector $\LQ$, respectively~(for $\lambda_{eff}=e$),
and $\LQ$ single production contributes  30-50\% to the total signal rate.

In this work,
the most general form of scalar and vector $\LQ$ interactions 
with quarks and gluons  has been implemented into CalcHEP/CompHEP packages,
which was one of the primary  aspects of this study.

\section*{Acknowledgments}
A.B.    grateful to  Oscar Eboli, C.-P. Yuan, Jon Pumplin, Thomas Nunnemann and John R. Smith 
for stimulating  discussions. C.L. and R.M. thank NSERC/Canada for their support.
DOE support is also acknowledged.
This work has been performed within the ATLAS Collaboration with the help
of the simulation framework and tools which are the result of the
collaboration-wide efforts.

\end{document}